\documentstyle[epsf,preprint,aps]{revtex}
\tightenlines
\begin{document}
\draft

\title{Properties of few-electron artificial atoms}

\author{
K. Varga$^{1}$\footnote{Corresponding author (e-mail: vargak@ornl.gov).},
P. Navratil$^2$\footnote{On leave of absence from Nuclear Physics Institute,
Academy of Sciences of the Czech Republic, 250 68 Rez near Prague,
Czech Republic}, J. Usukura$^3$  
and Y. Suzuki$^4$}
\address{
$^1$ Solid State Division, Oak Ridge National Laboratory, Oak Ridge,
37831 Tennessee, USA
\\
and
\\ 
Institute for Nuclear Research of the Hungarian Academy 
of Sciences (ATOMKI), 4000 Debrecen, Hungary PO BOX 51
\\
$^2$ Lawrence Livermore National Laboratory, P. O. Box 808,
Livermore, CA 94551 
\\
$^3$ Graduate School of Science and Technology, Niigata University, 
Niigata 950-2181, Japan
\\
$^4$ Department of Physics, Niigata University, Niigata 950-2181, Japan}

\maketitle
\vspace{1cm}

\begin{abstract}
The spectra of quantum dots of different geometry (``quantum ring'',
``quantum cylinder'', ``spherical square-well'' and ``parabolic'' )
are studied. The stochastic variational method on correlated 
Gaussian basis functions and a large scale 
shell-model approach have been used to investigate these ``artificial'' 
atoms and their properties in magnetic field. Accurate numerical results 
are presented for $N$=2-8 electron systems. 
\end{abstract}
\narrowtext
\section{Introduction}
The possibility of fabrication of artificial atoms or quantum dots 
with ``tunable''
properties is a fascinating new development in nanotechnology.  These
quantum dots not only offer the opportunity of  various applications 
(laser and electronic devices, memories, quantum gates, etc.) but 
they are quite intriguing physical systems in their own right. 
\par\indent
There are two very common ways to fabricate quantum dots. In the first 
method photolithography is used to create nanoscale electrodes on the 
surface of heterostructures and the confining potential is due to the 
electric voltage between these electrodes \cite{ashoori}.  
The second method uses material growth techniques to fabricate 
self-assembled quantum dots \cite{grundmann}. 
\par\indent
Most of the theoretical model calculations use the effective-mass
approximation to study the energy levels or other properties of the
electrons confined in quantum dots. These calculations address the 
low-energy sector 
where the interband mixing is assumed to be negligible and the periodic 
crystal potential is taken into account through the effective mass and
dielectric constant. In these models the electrons move in an external
confining potential and interact via the Coulomb interaction.  
\par\indent
Given the geometry of the quantum dot and the parameters of the 
heterostructure, the confining potential can be determined by a 
self-consistent calculation. This is not, however, a trivial task and most
work on quantum dots uses simple model potentials. The confinement
is generally very strong in the vertical $z$ direction creating quasi 
two-dimensional (2D) systems. The confinement on the $xy$ plane is most 
often assumed to be parabolic. A study of realistic confining 
potentials found that this approximation is fairly good in certain cases
but generally the confining potential might significantly differ
from a harmonic-oscillator one \cite{maksym}. 
\par\indent
The apparent similarity of ``natural'' atoms and quantum dots suggests the 
application of sophisticated theoretical methods used in atomic physics 
and quantum chemistry to calculate the properties of quantum dots.
Parabolically confined 2D quantum dots have been studied by several 
different well-established methods: Exact diagonalization techniques 
\cite{diag,ed},
Hartree-Fock approximations\cite{Fuj96,Mul96,Yan99}, 
and density functional approaches
\cite{Kos97,Hir99}. Quantum Monte Carlo (QMC) techniques have also been used
for 2D \cite{Bol96,Har99,umr,dean,egger}  as well as three-dimensional 
(3D) structures 
\cite{Shu99}. Few-electron artificial atoms of spherical \cite{Szaf99} 
and cylindrical \cite{Szaf00} symmetry in 3D have also been 
investigated in variational and Hartree-Fock frameworks. The strongly 
correlated  low electronic density regime got much of attention
due to the intriguing possibility of the formation of Wigner molecules 
\cite{Yan99,wm1,wm2}. The novelty of this paper is that we use a correlated 
basis function which gives very accurate results for different few-electron
systems. The accuracy becomes important when one studies subtle properties
such as level orders, weakly bound states, etc. We investigate different
models of quantum dots suggested by various authors. 
\par\indent
In this paper the variational method has been used to solve
the few-electron Schr\"odinger-equation. Two different trial 
function sets have been applied.
\par\indent
In the first case the wave function is expanded in terms 
of harmonic-oscillator shell model (SM) basis states. This
basis forms a complete set and the energy
is obtained by diagonalizing the corresponding eigenvalue
problem. The only approximation is the truncation of the 
basis. The dimension of the harmonic-oscillator basis 
quickly increases with the number of single-particle states
included and even the powerful Lanczos method becomes 
unfeasible. The advantage of this approach is that it is
simple. Once the matrix elements are calculated there is no need
for optimization of the basis set. 
In addition, we may improve on this approach
by utilizing the starting-energy independent two-body effective
interaction \cite{navr1} that takes into account two-electron correlations
from the excluded space. Again, no additional optimization is needed
as the effective interaction does not depend on any extra parameter.
\par\indent
In the second approach a Gaussian basis is used. This basis
is nonorthogonal and overcomplete. The trial function depends 
on the parameters of the Gaussians and one has to optimize
the parameters to get the best energy. The most adequate 
basis functions are selected by the stochastic variational 
method (SVM) \cite{book,prc}. The advantage of this basis lies in its 
flexibility. A relatively small number of basis functions 
give very accurate results
provided that the parameters are carefully optimized. 
\par\indent
We have carefully compared the results obtained by these
basis states to test the accuracy of the energies and other 
physical properties. Despite of the fact that several 
calculations exist for the 2D case, only Ref. \cite{umr}
reports numerical values of energy to the best of our knowledge. 
These quantum mechanical systems provide us with 
very good tests of different approaches, and therefore we think
that it is important and useful to tabulate the 
energy and other quantities of the quantum dots. This
may serve as a benchmark test to compare different methods. 
\par\indent
An intriguing 
feature of these systems is that the strength (and shape) of the confining 
potential can be changed. Unlike the natural atoms the relative importance
of the pairwise Coulomb interaction and the external potential can be tuned.
\par\indent
We have  calculated the ground and the first few excited states.
The ground and excited states are characterized by the total
orbital angular momentum $L$ and spin $S$. The order of the levels depends
on the external (confining) potential. We have investigated how the
level order changes as the parameter of the potential is varied. 
\par\indent
The next section introduces the basics of our formalism. The results
of the calculation for different systems are presented in sec. III.
The last section is devoted to discussion and summary.

\section{The formalism}
We investigate a system of ${N_e}$ electrons confined by the potential 
$V_{\rm con}({\bf r})$. 
The  Hamiltonian is
\begin{equation}
H=\sum_{i=1}^{N_e}
\left(-{\hbar^2\over2 m^*}\nabla_i^2+ V_{\rm con}({\bf r}_i)\right) +
{e^2\over\epsilon}\sum_{i<j}^{N_e}{1\over \vert{\bf r}_i-{\bf r}_j\vert}~~.
\label{eq1}
\end{equation}
In Eq. (\ref{eq1}), $m^*$ is the effective mass of the electron,
and $\epsilon$ is the dielectric constant of the semiconductor.
In  the following (if not explicitly specified otherwise) we
will use effective atomic units, defined by $\hbar=e^2/\epsilon=m^*=1$. In this
system of units, the length unit is the Bohr radius $(a=\hbar^2/m_ee^2)$
times $\epsilon/(m^*/m_e)$,
and the energy unit is the Hartree $(H=m_ee^4/\hbar^2)$
times $(m^*/m_e)/\epsilon^2$ where $m_e$ is the mass of the electron.
For the GaAs dots we consider here, $\epsilon=12.4$ and $m^*=0.067 m_e$,
and the effective Bohr radius $a_0^*$ and the effective Hartree $H^*$ are
$\simeq 97.94 \,{\rm \AA}$ and $\simeq 11.86$ meV, 
respectively. These effective 
length and energy units are called atomic units (a.u.) in what follows. 
\par\indent
In the variational method the trial wave function is expanded in terms of 
basis functions:
\begin{equation}
\Psi=\sum_{i} c_i \Phi_i,
\end{equation}
and the variational energies are obtained by solving the generalized 
eigenvalue problem
\begin{equation}
\sum_{j}(H_{ij}-EO_{ij})c_j=0,  \ \ \ \ \  
H_{ij}=\langle \Phi_i \vert H \vert \Phi_j \rangle
\ \ \ {\rm and} \ \ \ 
O_{ij}=\langle \Phi_i \vert \Phi_j \rangle.
\end{equation}
The energy eigenvalues $E_1, E_2, ... $ are variational upper bounds 
of the energies of the ground and first, second,.... excited states. 

\subsection{Correlated Gaussian basis functions}
The correlated Gaussian basis is defined in the following way:
\begin{equation}
\Phi_i({\bf r})={\cal A} \left\lbrace{\rm exp}(-{1\over 2}{\bf r} A_i {\bf r})
\theta_{LM_L}({\bf r}) \chi_{SM_S}\right\rbrace ,
\label{basisfnc}
\end{equation}
where ${\cal A}$ is the antisymmetrizing operator for the electrons and 
${\bf r}=({\bf r}_1,...,{\bf r}_{N_e})$ stands for a set of 
spatial coordinates of the electrons.   
${\bf r} A_i {\bf r}$ is a short-hand notation of the 
quadratic form $\sum_{j,k=1}^{{N_e}} (A_{i})_{jk} {\bf r}_j \cdot{\bf r}_k$, 
where $A_i$ is an $N_e\times N_e$ symmetric positive-definite 
matrix whose elements 
are variational parameters.  
Both  the spin function $\chi_{SM_S}$ and the angular function 
$\theta_{LM_L}({\bf r})$ are constructed by successively coupling
the corresponding single-particle functions:
\begin{equation}
\chi_{SM_S}=\left[\left[\left[\xi_{1\over 2}(1)\xi_{1\over 2}(2)
\right]_{s_{12}}\xi_{1\over 2}(3)\right]_{s_{123}}...\right]_{SM_S} 
\end{equation}
and
\begin{equation}
\theta_{LM_L}({\bf r})=
\left[\left[\left[{\cal Y}_{l_1}({\bf r}_1){\cal Y}_{l_2}({\bf r}_2)
\right]_{l_{12}}{\cal Y}_{l_3} ({\bf r}_3)\right]_{l_{123}}...\right]_{LM_L},
\end{equation}
where $\xi_{{1\over 2}m}$ and ${\cal Y}_{lm}({\bf r})=r^{l}
Y_{lm}(\hat{{\bf r}})$ are the spin and angular 
functions of the electron. 
\par\indent
The Hamiltonian we consider in 
this paper contains no term which couples the spin and 
orbital angular momentum, and commutes with the total spin and 
total orbital angular momentum or their $z$ components when 
the uniform magnetic field is applied in the $z$ direction. 
There is no coupling between the spin and the orbital  
part in the basis function of Eq. (\ref{basisfnc}).   
\par\indent
The correlated Gaussian function can be rewritten in a more 
intuitive form:
\begin{equation}
{\rm exp}\Big(-{1 \over 2} 
{\bf r}A{\bf r}\Big)=
{\rm exp}\left(-{1 \over 2} 
\sum_{k<l}^{N_e} \alpha_{kl} ({\bf r}_k-{\bf r}_l)^2 
-{1 \over 2} 
\sum_{k=1}^{N_e} \beta_{k} {\bf r}_k^2\right).
\end{equation}
$\alpha_{kl}$ and $\beta_k$ can be expressed by the elements of $A$ 
and vice versa. 
The advantage of this notation is that it explicitly connects
the nonlinear parameters $\alpha_{ij}$ to the pair correlation
between the particles $i$ and $j$ and thus explains the name ``correlated
Gaussians''. The second part, ${\rm exp}(-{1 \over 2} 
\sum_{k=1}^{N_e} \beta_{k} {\bf r}_k^2)$, is a product of independent
single-particle Gaussians. 

\subsection{Separation of the relative and the center-of-mass motion}
\label{separation}
When a system is subjected to an external field, 
its relative and center-of-mass motion 
cannot be separated. The harmonic-oscillator 
confinement is the only exception. To separate the relative
and center-of-mass motion one can introduce a relative (e.g. Jacobi) 
coordinate system $({\bf x}_1,...,{\bf x}_{N_e})$ and rewrite the Hamiltonian
and the wave function in terms of the relative coordinates.
If the Hamiltonian is translational invariant, then the center-of-mass 
Hamiltonian 
$H_{\rm cm}=-\frac{\hbar^2}{2M}\nabla_{\bf R}^2+\frac{1}{2}M\omega^2 R^2$
($M=\sum_{i=1}^{N_e}m^*$ and ${\bf R}={\bf x}_{N_e}$ 
is the center-of-mass coordinate) can be separated. 
The eigenenergies of the center-of-mass
Hamiltonian are $E_{\nu\lambda}=(2\nu+\lambda+3/2)\hbar \omega$.
In the following we are interested in the energies
$\epsilon_{\rm int}$ of the $H_{\rm int}=H-H_{\rm cm}$ internal Hamiltonian.
The lowest energy states of the system are given by 
$\epsilon_{\rm int}+E_{00}$. 
\par\indent
If the center-of-mass and the relative motion can be separated, then 
we use Eq. (7) by setting $\beta_k=0$ and the angular part (Eq. (6)) 
is replaced by a similar expansion, but now the arguments of the 
spherical part are the relative coordinates ${\bf x}_i$. 
In that case we have only $N_e-1$ independent variables.
\par\indent
For the pure harmonic-oscillator confinement case one can use both the
relative (the center-of-mass separated) and the single-particle 
coordinate (the center-of-mass motion included in the Hamiltonian
and the wave function) approach. The energy converges to the same 
value (except for the trivial $E_{00}$ shift). The solution in the relative
coordinate approach is, however, much easier as convergence is much
faster because the center-of-mass degrees of freedom is decoupled.

\subsection{Stochastic variational method}
The energy crucially depends on the variational parameters. The optimal 
nonlinear parameters are selected by the stochastic variational method
\cite{book,prc}. In each step of this procedure, ${\cal K}$ different $A_i$ are 
generated by randomly choosing the values of $\alpha_{kl}$ and $\beta_k$ 
from the 
$[0,\beta]$ interval. The parameter set which gives the best variational 
energy is selected and the function corresponding to 
that parameter set is added to the set of basis functions. 
The trial function also depends 
on the intermediate coupling quantum numbers ($s_{12},s_{123},...$) and 
$(l_1, l_2, l_{12}, l_3,...)$. 
These possibilities are also randomly tested during the optimization of
the basis. 
\par\indent
Our stochastic selection procedure uses the following steps:
\par\indent
(1) To set up a new basis or enlarge an existing one:
\par\noindent
Let us assume that the basis set has ${\cal N}-1$ elements.
One generates ${\cal K}$ random basis states and calculates the
energies $E_{{\cal N}i}$
($i=1,{\ldots}, {\cal K}$) with the
new ${\cal N}$-dimensional bases which contain the $i$th random
element and the preselected ${\cal N}-1$ basis elements.
The random state which
gives the lowest energy is selected as a new basis state and
added to the basis. The variational principle ensures that the
energy of the ${\cal N}$-dimensional basis is always lower than
that of the ${\cal N}-$1 dimensional one. This procedure therefore
guarantees to lead to a better and better upper bound of the
ground state energy. Notice that as the ${\cal N}-1$ dimensional basis
is orthogonalized this method does not require the diagonalization
of ${\cal N}$-dimensional matrices \cite{book,prc}. The energy
gain, $\epsilon_{\cal N}=E_{\cal N}-E_{{\cal N}-1}$, shows the rate of
convergence. A calculation of good convergence 
gives $\epsilon_{\cal N}\approx 0$.
\par\indent
(2) Refinement: To improve the energy of a basis:
\par\noindent
In the previous step only the newly added element is optimized, but the rest
of the basis is kept fixed. In the refinement we keep the 
dimension of the basis
fixed and try to replace the $k$th basis element with ${\cal K}$ randomly
generated elements. If the best energy obtained by substituting the
$k$th basis state with the random candidate is lower than that of the
original basis, then the $k$th basis state is discarded and the new
random state is included in the basis. This procedure is cyclicly
repeated for $k=1,{\ldots} ,{\cal N}$. As the dimension of the
model space is fixed, this step does not necessarily give lower energy, but
in practice in most cases it does. Actually if one cannot find better
basis elements, that is an indication of a well converged energy/basis.
Again no diagonalization is needed in this step when starting 
from an orthogonalized basis.
\par\indent
(3) Optimization by ``fine tuning'' of the parameters: 
\par\noindent
In step (2) the parameters are randomly selected  irrespective of
their previous values. This certainly helps to avoid the traps 
of local minima, but if one is already (presumably) close to
the ``global'' minimum then the chance to move closer to it is
small. If the basis parameters are ``reasonably'' optimized or 
further repetition of step (1) or (2) does not lead to appreciable
changes, one may try to change the basis parameters by selecting
new parameters in the vicinity of the existing ones. That increases the
probability of finding the nearby minimum. In practical calculations
this step was implemented by requiring the new random parameters 
to be in the $[0.8\alpha,1.2\alpha]$ interval ($\alpha$ is the 
previously chosen parameter). In this case the basis optimization is done in exactly 
the same way as in step (2). The only difference is that the search
interval is limited and defined by the previous parameters.
\par\indent
A combination of steps (1)-(2)-(3) is repeated until the required
accuracy is reached. A practical and economical way  to set up  a basis
is to generate  ${\cal N}$ elements (${\cal N}=20$ or 40 
is a reasonable choice) by using
step (1). Then repeat step (2) for each basis state several (say 3-5)
times. Use step (1) once more to enlarge the basis by adding ${\cal N}$ 
elements to it and repeat step (2) as described before. After 
reaching a certain basis
size where further repetition of steps (1) and (2) does not yield 
considerable improvement then try step (3). 
\par\indent
This basis selection procedure proved to be quite reliable and provides
a very accurate solution. More details can be found in \cite{book}.

\subsection{Harmonic-oscillator basis}
Alternatively, we also set up a harmonic-oscillator basis 
\cite{navr2}. In this case the basis functions are
\begin{equation}\label{sdbas}
\Phi_i({\bf r})={\rm det}\left\lbrace (\varphi_{n_j l_j m_j}({\bf r}_j) 
\xi_{{1\over 2} \mu_j}(j)) \right\rbrace,
\end{equation}
where the single-particle function $\varphi_{n_j l_j m_j}$ is a 
harmonic-oscillator function. This basis 
depends on only one parameter, the harmonic-oscillator 
width. For harmonic-oscillator confinement this is chosen 
to be equal to the oscillator frequency of the potential. In this way 
the harmonic-oscillator basis functions 
are eigenfunctions for a noninteracting system.
\par\indent
This is an orthogonal basis and the Hamiltonian matrix is sparse. 
The Lanczos method, in particular 
the Many-Fermion Dynamics shell-model code \cite{VZ94}, is used to find
the lowest eigenvalues. In the diagonalization we used all states up 
to $\sum_{i=1}^{N_e} (2n_i+l_i) \le N_{\rm max}$. 
\par\indent
The basic difference between the two bases is that the Gaussian basis 
is {\it explicitly} correlated. It explicitly depends on the 
$\vert {\bf r}_i -{\bf r}_j \vert$ distances, so it is better suited to 
describe the electron-electron correlations. At the same time the 
harmonic-oscillator basis is simpler because no optimization is needed. 
\par\indent
An advantage of the harmonic-oscillator basis is the fact that we may
alternatively perform the calculations 
in the Jacobi coordinates with the
center-of-mass degrees of freedom removed. It is straightforward, although
numerically intensive, to construct an antisymmetrized harmonic-oscillator
basis depending on the Jacobi coordinates \cite{navr1}. Depending
on the problem, we may choose the more efficient basis. For 
$N_e=3,4,5$ electron
systems it turns out that the use of Jacobi coordinates is more profitable.
For larger number of electrons, it is more efficient to use the single-particle
coordinates and the Slater determinant basis (\ref{sdbas}).  
\par\indent
As the harmonic-oscillator frequency is fixed as described above, the
only parameter of the calculation is the model space size characterized 
by $N_{\rm max}$. In the present calculations we use as large $N_{\rm max}$
as possible, typically  $N_{\rm max}=15-33$ for
$N_e< 5 $ and $N_{\rm max}=8-12$ for larger systems.

\par\indent
A speed up of convergence can be achieved by utilizing the effective
interaction approach that was succesfully applied in the {\it ab initio}
shell model calculations for few-nucleon systems and light nuclei 
\cite{navr1,navr2}.
While it is crucial in the nuclear physics application to use the effective
interactions, in the present electron systems the effective interaction
provides only minor improvement. In some cases, however, it brings
the SVM and the SM results to much closer agreement. The details of
how the effective interaction is computed from the bare Hamiltonian,
here the harmonic-oscillator and Coulomb interaction, is given, e.g., 
in Refs. \cite{navr1,navr2}. The basic goal of the effective interaction
is to take into account, in this case two-electron, correlations
from the excluded space, i.e., from the space containing excitations
above $N_{\rm max}$. 
A formal difference from the nuclear case is
that here the harmonic-oscillator potential is a real binding potential,
while in the nuclear application it is a model potential representing
the mean field formed by all nucleons, which is added/subtracted to the
real nucleon-nucleon interaction in order to facilitate the 
effective interaction calculation.    
\par\indent
We note that when the $m$-scheme basis (\ref{sdbas}) is used 
the good quantum numbers are checked by evaluating the mean values of 
relevant operators, e.g., $J^2$, $\bf L$, and $\bf S$ for each eigenstate.

\subsection{Magnetic field}
In external magnetic field the kinetic energy operator
is replaced by
\begin{equation}
{1\over 2m^*}
{\bf p}_i^2 \ \rightarrow 
{1\over 2m^*}({\bf p}_i+{e\over c} {\bf A}_i)^2 .
\end{equation}
We consider a uniform magnetic field ${\bf B}=(0,0,-B)$. 
By taking ${\bf A}_i=-{1\over 2} {\bf r}_i\times {\bf B}$ 
the above expression can be rewritten in a more detailed form 
\begin{equation}
{1\over 2m^*}({\bf p}_i+{e\over c} {\bf A}_i)^2=
-{1\over 2m^*}\hbar^2 \Delta_i
+{1\over 2} m^*(\omega_c/2)^2 (x_i^2+y_i^2)
-{1\over 2} \omega_c l_{zi},
\end{equation}
where $l_{zi}$ is the $z$ component of the orbital angular momentum of
the $i$th electron. The cyclotron frequency for the parameters
we use in this paper reads as
\begin{equation}
\hbar\omega_c={e\hbar B \over m^*c}=
{2m_e\over m^*}{\mu_B B}=0.14572 B \ (H^*),
\end{equation}
where the Bohr magneton is $\mu_B=e\hbar/(2m_ec)=0.05788$ meV/T. The
interaction of the magnetic field with the spins leads to the Zeeman term, 
$-g^*\mu_B B s_{zi}$, where $s_{zi}$ is the $z$ component of the
spin of the $i$th electron and $g^*$ is the effective $g$-factor 
of the electron. The
Zeeman term leads to the splitting of the energies for different spin
orientations. As the Hamiltonian with this term 
still commutes with the $z$ component of the total spin, 
$S_z=\sum_{i=1}^{N_e} s_{zi}$, the energy shift is simply given by 
$-g^*\mu_B B S_{z}$ and one can easily add this to the energies 
presented in the following. This energy is not included in what follows. 
\par\indent
The correlated Gaussians defined above are
not ideally suited for systems in magnetic
field, because the basis functions belonging to different orbital angular
momenta would be coupled by the Hamiltonian. This coupling would require
an infinite series of orbital angular momentum states, which is obviously 
out of question.
To avoid this, we choose a deformed form of the correlated Gaussians (DCG)
\cite{kvarga}:
\begin{equation}
{\rm exp}\left\lbrace 
-{1\over 2} \sum_{i,j=1}^{N_e} 
A_{ij} {\bf {\boldmath \rho}}_i \cdot {\bf {\boldmath \rho}}_j 
-{1\over 2} \sum_{i,j=1}^{N_e}
B_{ij} z_i  z_j 
\right\rbrace,
\end{equation}
where the nonlinear parameters are different (and independent)
in the $xy$ and $z$ directions (${\bf {\boldmath \rho}}_i=(x_i,y_i)$).
This extension brings a great deal of flexibility by allowing
a separate description on the $xy$ plane and along the $z$ axis.
The Hamiltonian does not commute with $L^2$ but it does
with $L_z$.  The eigenfunctions have good quantum number $M$ of $L_z$.
Note that we will use $M$ for the orbital angular momentum 
quantum number in 2D and $L$ for the one in 3D.
The above form of the DCG belongs to $M=0$. To allow for 
$M\ne0$ states we multiply the basis by\cite{book}
\begin{equation}
\prod_{i=1}^{N_e} \xi_{m_i}({\bf {\boldmath \rho}}_i),
\end{equation}
where
\begin{equation}
\xi_{m}({\bf {\boldmath \rho}})=
\cases {
(x+iy)^m & for nonnegative integer $m$ \cr
(x-iy)^{-m} & for negative integer $m$.} 
\end{equation}
Thus our variational basis function reads as
\begin{equation}
\Phi_M({\bf r})=
{\cal A} \left\lbrace
\Bigg(
\prod_{i=1}^{N_e} \xi_{m_i}({\bf {\boldmath \rho}}_i)
\Bigg)
{\rm exp}\Bigg(
-{1\over 2} \sum_{i,j=1}^{N_e} 
A_{ij} {\bf {\boldmath \rho}}_i \cdot {\bf {\boldmath \rho}}_j 
-{1\over 2} \sum_{i,j=1}^{N_e} 
B_{ij} z_i z_j 
\Bigg)
\right\rbrace ,
\end{equation}
where $M=m_1+m_2+...+m_{N_e}$.
\par\indent
The above basis is defined for 3D cases. It is used not only for 
solutions in the presence of magnetic field but also for external 
potentials with cylindrical symmetry. For 2D calculations the same 
form is used except that the 
third component of the vectors are dropped (or equivalently
$B_{ij}=0$ is assumed). 

\section{Calculation}

\subsection{Harmonic-oscillator confinement in 2D}
The harmonically confined 2D systems received much theoretical
attention and this is a very good test case to gauge the accuracy
of different approaches. In this case the confining interaction takes the
simple $V_{\rm con}(r)={1\over 2} m^* \omega^2 r^2$ form. 
The single-particle energy of the harmonic-oscillator potential 
is given by $(2n+|m|+1)\hbar \omega$, where $n=0,1,2,\ldots$, and 
$m=0,\pm 1, \pm 2,\ldots$. In Table I we compare our results to 
the ``exact diagonalization'' \cite{diag} and the QMC 
methods \cite{Bol96,Har99,umr} for the $N_e=3$ electron system. We have 
carefully optimized the parameters and repeated the calculation several 
times to check the convergence. Our result is expected to be accurate up 
to the digits shown in Table I. In principle the QMC calculations, 
except for the statistical error, give the exact energy of the system. 
In practical cases the famous ``minus-sign problem'' forces 
the QMC approaches to use certain approximations (in 
Ref. \cite{umr,Har99} the ``fixed-node'' method has been used).
The slight difference between our results and the QMC values is
probably due to this fact.
The energies for both the ground and excited states are in good 
agreement. Our results
are slightly better than the other calculations in each case. 
\par\indent
In Table II a similar comparison is presented for $N_e=2-6$ electron 
systems. The QMC results \cite{umr} quoted in Table II are obtained by 
very careful calculations and their statistical error is very small.
Note that the confining strength is slightly different in the calculations
presented in Tables I and  II. This table also includes 
the virial factor 
\begin{equation}
\eta=2\langle T \rangle /\langle W \rangle, \ \ \ \ 
\langle W \rangle =\langle \sum_{i=1}^{N_e} 
{\bf r}_i \cdot \nabla_i V_{\rm int} \rangle, 
\end{equation}
where 
$V_{\rm int}$ is the ``interaction part''
of the Hamiltonian, including the confining and the electron-electron 
interactions. The virial factor is unity 
for the exact wave function. 

\par\indent
Our result is in excellent agreement with
the QMC predictions \cite{umr} in all but one case ($N_e=4$).
The QMC renders the
$(M,S)=(0,0)$ state to be the ground state and the $(M,S)=(0,1)$ state 
to be the first excited state, 
which is a violation of the Hund's rule. The shell filling and Hund's 
rule has been experimentally investigated in Ref. \cite{taru} and it is
found that a circular dot obeys the Hund's rule. According to 
the Hund's rule the ground state of a system with a well developed shell 
structure is in the maximum spin state allowed by the Pauli principle. 
See the Appendix for an example of the  $N_e=4$ electron case. 
The violation of the Hund's rule in this system is 
also observed in  another QMC calculation \cite{Bol96}. This latter 
calculation predicts a relatively large energy difference
between the (0,0) and (0,1) levels, but it is somewhat less accurate 
for $N_e=4$, using only the lowest Landau levels. 
\par\indent
Our calculations, in agreement with the Hund's rule, predicts the
$(M,S)=(0,1)$ state to be the ground state and the (2,0) and (0,0)
states to be the lower excited states.
This contradicts the results of \cite{Bol96,umr} 
but is in agreement with the other QMC calculation\cite{Har99}.
Our other energies are very close to the QMC results: The agreements 
for $N_e=5$ and $N_e=6$ electron systems are very impressive. 
\par\indent
We define the pair correlation function 
\begin{equation}
P({\bf r}, {\bf r}_0)={2\over N_e(N_e-1)}
\langle \Psi |\sum_{i<j}
\delta({\bf r}_i-{\bf R}-{\bf r})
\delta({\bf r}_j-{\bf R}-{\bf r}_0)| \Psi \rangle.
\end{equation}
Here ${\bf r}_0$ is a fixed vector and its magnitude is chosen to be 
equal to $\langle \Psi |\sum_i |{\bf r}_i-{\bf R}||\Psi \rangle
/N_e$. The function $P({\bf r}, {\bf r}_0)$ gives us information 
on where one electron located at ${\bf r}_0$ sees other electrons. 
Figures 1 and 2 display the pair correlation functions for the ground 
state $(M,S)=(1,1/2)$ and the first excited state $(M,S)=(2,3/2)$ 
of $N_e=5$ electron system. Both figures show qualitatively similar 
features. 
For $\omega=1$, the confinement potential 
is strong and the contribution of the single-particle energies 
to the total energy is larger than that of the Coulomb potential. 
The electrons are confined in a rather compact region so that the 
contour map does not show clear four peaks. On the contrary, for 
$\omega=0.1$ the effect of the confinement becomes weak and the 
contribution of the Coulomb potential is larger than that of 
the harmonic-oscillator part. The size of the system grows and 
we see clearly well-separated pentagon-like structure. 
\par\indent
Next we present in Table III an example where the magnetic field 
is nonzero. Again, the energies are in good agreement with the 
QMC \cite{Bol96} and diagonalization \cite{diag} methods.
In 2D the inclusion of the magnetic field leads to a change of
the harmonic-oscillator frequency 
\begin{equation}
\omega \rightarrow \sqrt{\omega^2+(\omega_c/2)^2}
\end{equation}
and an energy shift by $-{1\over 2} M\hbar\omega_c$, so we expect 
that our results are as accurate as those for the zero field case. 
The accuracy is also indicated by the virial factor included in Table III.  
\par\indent
We have improved the prediction of the diagonalization method 
\cite{diag}. The diagonalization method would give the ``exact'' solution
in infinite model space. In practice the diagonalization is always limited
to finite dimensions. The slight disagreement between our and QMC 
results might be due to the statistical (and/or fixed node) error 
of the QMC calculation. 
\par\indent
The states listed in Tables I and  III 
(i.e., (1,1/2), (2,1/2), (3,3/2)) are quoted as the lowest-energy
states in several papers\cite{diag,Bol96,Har99}. Figure 3 shows that this
is not the case. The level order at $B=0$ is
$(1,1/2),(0,3/2),(2,1/2),(0,1/2),(3,3/2)$.
Figure 3 shows that the spin unpolarized $(1,1/2)$ state is the ground state
in the weak magnetic field limit and the $(2,1/2)$ unpolarized state becomes
the ground state in a very small interval of the magnetic field strength. 
The spin polarized states become the ground state above $B=$2.5 T. The figure 
also shows that
the lowest spin polarized state is the $(0,3/2)$ state for weak field. For
stronger field the $(3,3/2)$ and then the $(6,3/2)$ states become the lowest
spin polarized (and ground) state, following the $(3,6,..., 3n)$ ``magic''
sequence. Other spin polarized states (e.g., (1,3/2) etc.) never become
the lowest state. The explanation of the magic sequence is very simple.
In the spin polarized case all electrons have to occupy different orbits. As
the magnetic field gets stronger, the single-particle states belonging to
positive orbital angular momentum quantum numbers ($m_i=0,1,2,3,...$) are
energetically more favorable than those with negative ones. The $M=3$ state
($(m_1,m_2,m_3)=(0,1,2)$) is therefore lower than the  $M=2$ state 
(which requires $(0,-1,3)$
or $(1,2,-1)$, etc.). For the weak magnetic field the above 
argument does not hold in general
and the lowest polarized state is $M=0$ with the $(0,1,-1)$ orbits. 
\par\indent
Similar picture is valid for $N_e=4$ (see Fig. 4). In the very weak field
regime the unpolarized $(M,S)=(0,1)$ state is the ground state. By
increasing the magnetic field, the spin polarized $M=2$ state 
($(m_1,m_2,m_3,m_4)=
(0,1,-1,2)$) becomes the ground state before the ``magic'' $M=6 \, 
(0,1,2,3)$ 
state takes over. 
\par\indent
Figure 4 reassures that the $(M,S)=(0,1)$ state is the ground state and the 
$(M,S)=(0,0)$ is an excited state for zero magnetic field: Both states
belong to $M=0$, and therefore the change of the magnetic field simply 
changes the harmonic-oscillator frequency (see. Eq. (18)). The figure thus 
shows that the
order of these two states remains the same for different harmonic-oscillator
frequencies.

\subsection{Harmonic-oscillator confinement in 3D}
We have calculated the energies of the ground and first few excited states
of 3D few-electron systems confined by a harmonic-oscillator 
potential $(V_{\rm con}(r)={1\over 2} m^*\omega^2 r^2)$. 
The results 
for different values of the oscillator frequency are compared in Tables IV-X.
All intermediate spin coupling possibilities ($s_{12},s_{123},...$) 
are included in the trial function. The partial wave components 
($l_1,l_2,...)$ are included up to $\sum_{i=1}^{N_e-1} l_i \le 6$.

\par\indent
The two-electron case is relatively simple and it is analytically 
solvable for certain frequencies\cite{taut}. For $\omega=0.5$ for 
example, the exact energy is 2 a.u.\cite{taut} and we can easily 
reproduce this value up to several digits as shown in Table IV, where
the energies of other low-lying states are also listed. 
Three very different
oscillator frequencies are used to test the accuracy of the method under
different circumstances. In the case of $\omega=0.01$ the confinement 
is extremely weak and the Coulomb interaction governs the dynamics.
In the other limiting case the confinement is very strong $(\omega=10)$.
Another reason for choosing these values is that we want to study
the ordering of the energy levels as a function of the strength of the 
confining interaction. In the two-electron case, for example, 
there is a level crossing between the state $(L,S,\pi)=(3,1,-)$ and the
first excited state of $(0,0,+)$. The order of the other states listed
in Table IV does not change. 
\par\indent
The energies of the ground and excited states calculated by 
the correlated Gaussian and the harmonic-oscillator shell model 
basis are compared 
for  $N_e=3-6$ electron systems in Tables V-VIII. Both methods give very
similar results: The agreement is especially good for the ground and
first excited states. For higher excited states the Gaussian
basis gives slightly less accurate energies because it is significantly
more difficult to optimize the basis for excited states. 
In addition to the present results we note 
that the harmonic-oscillator calculation dependence on the model space size
for $N_e=3,4$ and $\omega=0.5$ was discussed and tabulated in Ref. 
\cite{navr3}.
\par\indent
The agreement is especially striking for $N_e=3$. Almost all digits are
equal for most of the calculated cases.
It is interesting to compare the order of the states in 2D and 3D.
In the 2D case for $N_e=3$ the energy levels of the first few states
follow the order of
$(M,S)=(1,1/2),(0,3/2),(2,1/2),(0,1/2)$, while  in 3D the levels
are ordered as $(L,S,\pi)=(1,1/2,-),(1,3/2,+),(2,1/2,+),(0,1/2,+)$.
This shows that
the lowest levels are built up from the same single-particle states:
In the 3D ground state  two electrons are in  the $l=0$ orbital
and one is in the $l=1$ orbital. The first excited state
has two electrons in the $l=1$ orbital, which are coupled to $L=1$
because their spin must be parallel to build up $S=3/2$ with the third.
In the 2D case they are in the 
$m=1$ and $m=-1$ orbitals and their total orbital angular momentum is 
$M=0$. 
The higher excited states have similar correspondence.
The same similarity occurs for $N_e=4$
($(M,S)=(0,1),(2,0),(0,0)$ in 2D and $(L,S,\pi)=(1,1,+),(2,0,+),(0,0,+)$
in 3D). For example, the 3D ground state has two electrons in
the $l=0$ and two electrons
(with parallel spin) in the $l=1$ orbital and the two electrons in the $l=1$
orbital are again coupled to $L=1$. In the 2D ground state
the two electrons are in the $m=1$ and $m=-1$ orbitals and the orbital
angular momentum is $M=0$.
With respect to the single-particle state 
occupations there is of course a big difference between
the 2D and 3D case. In 2D the shell fillings occur at $N_e=2,6,12,20,\ldots$
etc.,  while in 3D the shells are 
filled at $N_e=2,8,20,40,\ldots$ .
For $N_e>6$ particle systems the single-particle components of the wave
functions in 2D and 3D might be quite different. 
\par\indent
The addition energy is conveniently used to show the shell closure which 
occurs at a specific electron number. The addition energy $\Delta \mu(N_e)$ 
is defined by 
\begin{equation}
\Delta \mu(N_e)=\mu(N_e+1)-\mu(N_e),
\end{equation}
where the chemical potential 
$\mu(N_e)$ is the increase of the ground state energy 
by adding one electron to the ground state of $N_e-1$ system:
\begin{equation}
\mu(N_e)=E(N_e)-E(N_e-1).
\end{equation}
The shell or half shell closure is reflected by a sudden increase of 
$\Delta \mu(N_e)$ at a certain $N_e$ or the change of the 
differential capacitance given by $e^2/\Delta\mu(N_e)$. 
This is because the electron 
needs much energy when it fills an orbit across the degenerate orbits 
of a shell or goes beyond the half-shell due to the Hund's rule.  
The addition energies of the harmonically confined electrons
in 2D and 3D are compared in Figs. 5 and 6. In 2D the addition energy 
shows a large peak at $N_e=2$ and a smaller peak at $N_e=4$. The 
former corresponds to the filling of the $n=0, m=0$ orbit, while 
the latter is a reflection of the half shell filling of the degenerate 
orbits $n=0,m=\pm 1$, which can be understood by the Hund's rule. 
By decreasing $\omega$ the level spacing of the single-particle 
orbits becomes smaller and the correlation due to the 
Coulomb interaction takes over the shell structure. This explains 
why the peak at $N_e=2$ disappears for $\omega=0.1$. The behavior 
of the 3D addition energy is similar to the 2D case. One difference 
is that the half shell filling occurs at $N_e=5$ because 
the relevant orbit is $l=1$ and can accommodate six electrons. 

\par\indent
The results for $N_e=5$ and $N_e=6$ are somewhat less accurate 
and the agreement
between the SVM and the shell model is not as good as for $N_e<5$. The SVM
seems to be more accurate than the shell model for week confinement, 
where the role of the Coulomb interaction is more pronounced and
it is more difficult to take the Coulomb correlation into account with the 
shell model basis. At the same time it is more easy to use the shell
model approach for larger systems (see Tables IX and X), 
while the SVM becomes very time consuming beyond $N_e=6$. 
\par\indent
A general feature of the results is that the excited states change their
level orders as the harmonic-oscillator strength changes, but the
ground state always remains the same. We have very carefully tested this 
property and we do not find any level crossings with the ground state. 
\par\indent
Other insights of the relation between the 2D and 3D systems can be 
gained by comparing the expectation values of the kinetic, confining and
Coulomb operators. 
Tables XI and XII show the contribution of 
the Coulomb, kinetic and confinement parts of the Hamiltonian 
to the total energy.
The contributions are nearly equal in the $\omega=0.5$ 
case. Just as one expects in the strong confinement case 
($\omega=10$) the kinetic and confinement energies are 
strongly enhanced, and the Coulomb energy is relatively small
but not negligable. On the other hand, in the weak confining
case the Coulomb interaction dominates.
\par\indent
The contribution of the confining interaction, and thus the kinetic
energy is of course larger in 3D. If the electrons would not interact 
then both the kinetic and the harmonic confinement energies would be 1.5
times increased in 3D compared to the 2D case. In the interacting case the 
kinetic and confinement energy increase is roughly 1.5 for $N_e=2$ and 
$N_e=3$.
For larger systems the increase is smaller. On the other hand, the 
Coulomb correlation energy is smaller in 3D than in 2D because there is 
larger space available in 3D for the electrons. 

\subsection{Spherical square well}
As an alternative to the harmonic confinement one can consider a spherical
square well model of the 3D quantum dots. In this case the electrons 
are confined by a square well potential:
\begin{equation}
V_{\rm con}(r)= 
\cases{
-V_0 &  $r \le R$ \cr
 0 &  $r > R$ . }
\end{equation}
The  square well potential is analytically solvable for
one particle case. The eigenenergies $E$ can be determined from
the transcendental equation, 
$\sqrt{V_0-|E|} {\rm cot} \big(\sqrt{2(V_0-|E|)} R\big) = - \sqrt{|E|}$
(for $l=0$, and atomic unit used). Our SVM numerical 
approach virtually exactly reproduces the analytically determined energies. 
\par\indent
Spherical quantum well-like quantum dots have been studied in 
Ref. \cite{Szaf99}.
Unlike the harmonic-oscillator potential, 
the spherical well can only hold a certain
number of electrons. The number of electrons that a spherical well can bind
depends on  $V_0 R^2$. Figure 7 shows the energies of few-electron
systems confined by a spherical square-well potential in 3D as a function of
the radius $R$. A spherical well can only bind an electron if
$\pi^2/8 < V_0 R^2$. In our example $V_0=10$ and therefore the one electron
bound state appears when $0.35 < R$.
By increasing the radius  the two, three, ... etc.
electron systems may become bound in the well (see Fig. 7). 
This potential parameter has been used in Ref. \cite{Szaf99} to simulate
quantum dots in GaAs/Al$_{1-x}$Ga$_x$As with $x\approx 0.1$. Note that 
due to the difference in units the radius used here corresponds 
to $\sqrt{2}$ times the one used in Ref. \cite{Szaf99}. 
\par\indent
A comment is in order concerning the energy curves in Fig. 7 (and 
in Figs. 8 and 13 in later subsections). 
If the $N_e$ electron system has a bound ground ground state then our 
calculation converges to the energy of that state. If there is no bound
state in a given potential then the energy convergies to the lowest 
relevant threshold, which is in this case the energy of the $N_e-1$ 
electron system. In the figures the system is bound if the energy of the
$N_e$ electron system is below the corresponding $N_e-1$ electron system.
Strictly speaking, for unbound ($N_e$ electron) states the energy of 
the $N_e$ electron and $N_e-1$ electron systems should be equal. The 
convergence of the energy of the unbound $N_e$ electron system to the
energy of the $N_e-1$ electron system is rather slow, so one needs many 
basis states to describe the ``free'' electron. Therefore the fact 
that the energy curves of the unbound states are above the corresponding 
thresholds is the consequence of our spatially limited basis. By using 
more basis states and by allowing them to go far outside of the range 
of the confining interaction one would get the same energy for the $N_e-1$
and the unbound $N_e$ electron systems.
\par\indent
We have found no ``phase transition'' in $N_e=2$ and $N_e=3$ electron systems. 
The authors of Ref. \cite{Szaf99} have investigated
the energy of the lowest spin polarized and spin unpolarized $N_e=2$ and 
$N_e=3$
electron systems as a function of the radius of the square well. They found
that beyond a certain radius the spin polarized state becomes lower than
the spin unpolarized ground state. We have very carefully investigated
these systems and have not observed this ``para to ferromagnetic phase 
transition''. The same authors in  a later paper \cite{Ada00} 
investigated a harmonically confined two-electron system and found that the
spin unpolarized to spin polarized transition is most likely  an artifact of
the neglection of part of the electron-electron correlation in Hartree-Fock
calculations. 
\par\indent
We have increased the radius gradually (see Fig. 7).
As the Coulomb repulsion decreases the energy of the system gets smaller
and smaller, converging toward the energy of the noninteracting electrons
in the quantum well. If there is no Coulomb interaction then the energy
of the spin polarized and unpolarized electrons is the same, so by increasing
the radius both converge to the same energy. In our present example
($V_0=10$) the energies of the lowest lying spin polarized and 
unpolarized states are nearly degenerate beyond $R=15$, but we observe no
level crossing between them. 

\subsection{Cylindrical well: ``Quantum cylinder''}
In this section we present a calculation for cylindrical quantum dot.
Similar case has been considered in Ref. \cite{Szaf00} in an 
unrestricted Hartree-Fock framework. 
The confinement is defined as
\begin{equation}
V_{\rm con}({\bf r})=
\cases {
-V_0  & if   $(x^2+y^2)^{1/2} < R$  and $\vert z\vert < a$  \cr
 0    & otherwise. \cr
}
\end{equation}
In this case the spherical symmetry is broken, and only the $z$ component
of the orbital angular momentum is conserved. We have to use the 
DCG basis functions that were introduced for magnetic field.
\par\indent
First we consider a model potential with $V_0=10$ and $R=1$ and  change
the ``thickness'' of the dot (the height of the cylinder) from
$a=10$ to $a=0$ (in a.u.). In this way we transform the system
from a rod-like ($a=10$) geometry to a 2D disk ($a=0$).
Just like in the case of the spherical quantum well a quantum cylinder
can bind only certain number of electrons, depending on the potential
parameters $V_0$, $R$ and $a$.
The energy dependence on the thickness of the cylindrical dot
for $N_e=1-4$ electron systems is presented in Fig. 8. The figure shows
that as one expects the cylinder can hold more and more electron as
the size (in our case the height) increases. The real interesting thing
here is that the order of the energy levels also depends on the height
of the cylinder. For long, rod-like cylinders the ground state tends to
be the $M=0$ orbital angular momentum state. This probably means that the
electrons are equidistantly positioned along the $z$ axis. By decreasing
the height we approach a disk-like geometry which is somewhat similar
to the 2D harmonic confinement discussed earlier. And indeed, the level
order changes (see Fig. 8) and one has the same level order as in the
2D harmonic confinement case. In this way we have found an interesting
transition: By decreasing the height of the cylinder the $(0,1/2)$ and
$(1,1/2)$ $N_e=3$ electron  (also $(0,0)$ and $(0,1)$ $N_e=4$ electron) 
ground/excited states change their order. 
\par\indent
Figures 9 and 10 show the density of the $N_e=2$ electron system
in a ``long'' cylinder ($a=10$) along the radius and along 
the symmetry axis, 
respectively. The radial density distribution of 
the triplet $(1,1)$ and singlet $(0,0)$ states are very similar. 
Both peaks around 0.5 a.u. and the tail goes a little bit outside 
of the cylinder. 
The density of the triplet and the singlet state
in the $z$ direction, however, are very different as shown in Fig. 10. 
The cylinder is so long that the two electrons can be far away from 
each other to minimize the Coulomb repulsion and the density tail hardly goes
outside the cylinder.
\par\indent
Figures 11 and 12 show the density distribution for a ``thin''
cylinder ($a=1$). Decreasing the height of the cylinder the triplet state 
becomes less and less bound. 
The radial density distribution of this very weakly bound
triplet state has one peak in the 
cylinder and another outside. The density
distribution of the triplet pair along the $z$ axis has a peak inside the box
but also shows a long tail outside. So one of the electrons is inside the
box and the other one is already mostly outside. 
By decreasing the height further
the electron which is outside will be unbound, and 
the cylinder will be able to
bind only one electron. The singlet state is still bound for $a=1$ but 
it will also be unbound if one  decreases the height of the cylinder
further.

\subsection{``Quantum ring''}
Ring-like nanostructures have been grown by electron-beam lithography
\cite{ring1}. The electronic and magnetic properties of a single
electron quantum ring have been studied in Ref. \cite{ring2}.
We restrict our attention to a pure 2D case. An additional confining
interaction in the $z$ direction would cause no extra difficulty in  our
approach. 
The confinement in this case is defined as
\begin{equation}
V_{\rm con}(\rho)=
\cases{
0    &  $\rho < r_1$ \cr
-V_0 &  $r_1 \le \rho \le r_2$ \cr
0    &  $\rho > r_2 $.
}
\end{equation}
This describes a square well potential in a ring between
$r_1$ and $r_2$ on the $xy$ plane. 
\par\indent
The number of electrons bound in a ring-like potential, similarly 
to the previous cases, depends on the parameters ($V_0,r_1,r_2$) of 
the potential. An example of the energy levels in the model 
potential is presented in Fig. 13. The maximum number of electrons this
potential can bind is $N_e=4$. In the four electron case the lowest
state is the $(M,S)=(0,1)$ state just like in the case of 2D harmonic
confinement. In the present model potential the first excited state
$(0,0)$ is not bound with respect to the three-electron threshold.
\par\indent
The density distribution of an
electron in a quantum ring is shown in Fig. 14. The electron is
along the ring between $r_1$ and $r_2$. For the potential strength
used in this example the distribution is ``wider'' than the width
of the ring. By increasing the potential strength the ``inner'' and
``outer'' tails of the density are pulled more and more inside
the quantum well. 
\par\indent
This density distribution can be easily manipulated
by a perpendicular magnetic field. The magnetic field acts as a confining
harmonic-oscillator potential on the $xy$ plane. By increasing the
strenght of the magnetic field,  the density distribution
starts to move inward as shown in Fig. 15. 
In a certain very narrow region of the magnetic field
strength it has two peaks:  An outer peak centered in the ring and an
inner peak which is inside the harmonic confinement induced by the
magnetic field. If the magnetic field is stronger than a given value
then the electron moves inside of the harmonic confinement. This geometry
gives us a possibility of moving the electron from one well-defined
position to another one by switching on and off the magnetic field. Notice
that we have two sharply separated peak positions in this case. In the case of
the previously studied harmonic or spherical square well confinement
the density distribution also moves toward the origin in the magnetic field.
In that case, however, what we see is more like a ``shrinking'' of the
density distribution on the $xy$ plane. The change is more drastic in the case
of the quantum ring. The peak of the distribution shifts from one position to 
another. 

\section{Discussion}

The properties of artificial atoms created by confining electrons
in quantum dots of different geometry are qualitatively very similar.
The electrons occupy the single-particle orbits defined by the confining
interaction. The occupancy is determined by the Pauli principle and
the minimization of the Coulomb energy. In different confining potentials
the energy levels are different but the basic features are very similar. 
\par\indent
The confining interactions considered in this paper depend on one or
more parameters (harmonic-oscillator width, radius and strength of 
square-well, etc.). We have studied the dependence of the energy levels
on these parameters. One intriguing property that we have found is that in
spherically symmetric systems the ground state remains the same for 
any values of the parameters. Its energy level does not cross with that
of the excited states. At the same time, the order of energy levels of the
excited states frequently changes depending on the parameters of the
confining interactions. The change of symmetry of the ground state
of a spherical quantum well has been reported in Ref. \cite{Szaf99}. 
We have investigated few-electron systems in spherical quantum wells
of different parameters but we have not observed any similar change.
This confirms that such change of energy levels might be an artifact
of Hartree-Fock Calcualtion \cite{Ada00}.
Our calculation predicts that the ground state is in accordance with the
Hund's rule for any parameter values and there is no transition from
spin unpolarized to spin polarized states. 
\par\indent
The change of the ground state would give us an interesting possibility:
In a two electron system, for example, the ground state is a spin singlet, 
and the first excited state is a spin triplet. This two-state system may 
serve as a ``qubit'', an elementary gate for a quantum dot quantum computer. 
One would prepare a dot with singlet ground and triplet excited state
and a second one, with different geometry, where it is the other way around. 
The electrons can be moved from one dot to the other by an external 
electric field, for example, switching from $S=0$ to $S=1$. 
The calculations show, however, that 
no matter how do we change the geometry, the ground state does not change
for spherically symmetric systems. 
\par\indent
If the spherical symmetry is broken, for example by a magnetic field
or by a cylindrically symmetric confining potential, then the ground 
state and excited state energy levels  may cross each other. The fact that 
the magnetic field changes the order of energy levels has been studied 
in many works. In this paper we have presented a method which very accurately
predicts the level crossing as a function of the strength of the
magnetic field. 
\par\indent
The cylindrical quantum dot shows a very interesting example where the 
order of energy levels depends on the height of the quantum dot. In a 
disk-like cylinder the ground state obeys the Hund's rule, but for a
longer cylinder, typically when the height of the cylinder is 
larger than its diameter, the Hund's rule is violated. It would be interesting
to look for experimental 
evidence showing that the ground state of a cylindrical 
three- (four) electron quantum dot is 
a $(M,S)=(1,1/2)$ ($(M,S)=(0,1)$) state if the height of the cylinder 
is small and the ground state becomes a $(M,S)=(0,1/2)$ ($(M,S)=(0,0)$)  
state by increasing the height of the cylinder as predicted here. 
\par\indent
Harmonically confined electron systems in 2D have attracted enormous
attention. In this work we have also calculated 3D electron systems in
harmonic confinement. The comparison of the 2D and 3D cases shows the
effects of the quantum-well confinement in the $z$ direction in quantum dots.
For the same harmonic-oscillator strength the electrons are somewhat
farther from each other in 3D than in 2D, resulting in a smaller 
Coulomb energy in 3D. The energy difference between the 2D and 3D
geometry is dominantly due to the confinement and the kinetic energy. The 
qualitative features of the 2D and 3D systems are very similar in the case
of the few-electron systems investigated here. One can make an easy 
correspondence between the orbital and spin quantum numbers of the 
energy levels in 2D and 3D. The applicability of our method is tested
by calculations for  very different confining strength. The accuracy is
slightly worse for the weak confining region where one needs more
basis functions to achieve convergence.  But the accuracy is fairly good 
as one can judge it 
by the virial factor and by comparing with the results of 
different methods. In the weak confining region 
($\omega=0.01$, see Tables XI and XII) the contribution of the 
kinetic energy is fairly small compared to that of the Coulomb 
and  confining interactions. This suggests the existence of Wigner
crystal like structure both in 2D and 3D. 
Contrary to the prediction of Refs. \cite{Bol96,umr} we find that 
the ground state of $N_e=4$ system obeys the Hund's rule.
\par\indent
We have also investigated an example of a ring-like quantum dot in magnetic
field. This geometry offers an interesting possibility. In the case of zero 
magnetic field the electrons are distributed along the ring. By applying the 
magnetic field perpendicularly to the plane of the ring the electrons 
can be completely moved from the ring to the vicinity of the origin. 
Thus one may have the electrons in two very well separated regions.
\par\indent
The major difference between the harmonic and the square 
well confinements (cylindrical, spherical and ring) is that the harmonic
case can bind any number of electrons. The number of electrons bound in the
square-well case is finite and strongly depends on the parameters of
the potential. In that case one can predict how many electrons can be
bound in a certain quantum dot, that is, the ``capacity'' \cite{ada} 
of the dot. This is expected to be a more realistic model of quantum dots.
\par\indent
The concrete potential parameters and the potential itself can only be 
determined experimentally. In this work we have tried to follow the
``experimentally inspired'' and widely used potentials and parameters.
The aim of this work was to demonstrate the wide range of applicability
of the method and the investigation of different properties of various
artificial atoms. The direct comparison to experiments may not be 
suitable in this level. The general features of experimental findings
may be reasonably  well described by the potential models
considered here. In the present level one assumes a Hamiltonian which 
models the quantum dot and we try to solve this well-defined quantum 
mechanical problem in a careful manner. There are of course  
many things which may limit the applicability of our model 
Hamiltonian, but some of the phenomena that are experimentally observed can
be understood by such model calculations and hopefully some of the 
predictions of such models can be experimentally observed. 
\par\indent
The accuracy presented here is very useful and important  in the weakly 
confined (but strongly correlated) regime where otherwise it is 
difficult to predict the ground state, etc. One should also mention that
the comparison of various methods for these quantum mechanical problems
greatly helps the test and developement of different quantum mechanical 
many-body approaches. A nice example can be found in Ref. \cite{navr3}, 
where the solution of few-electron quantum dot problems helps to test the
Faddeev method which  was developed for nuclear few-body systems. 
\par\indent
In summary, we have presented a large scale variational approach to 
describe the spectra and other properties of artificial atoms. Different 
(parabolic, cylindrical, spherical and ring-like) confining interactions
have been investigated. The effects of magnetic field have also been 
studied. One of the aim of this paper was to introduce the method and
tests its capabilities on various models of quantum dots used in the
literature. Future work to investigate double quantum dots is under way. 

The work of K.V. is sponsored by the U.S. Department of Energy under 
contract DE-AC05-00OR22725 with the Oak Ridge National Laboratory, 
managed by UT-Battelle, LLC,  and OTKA grant No. T029003 (Hungary).
Y. S. is supported in part by the Matsuo Foundation and the 
JSPS-HAS cooperative research program. J. U. is supported 
by JSPS Research fellowship for Young Scientists. 
The work of P. N. was performed under the auspices of the U.S. Department
of Energy by the University of California Lawrence Livermore National
Laboratory under contract No. W-7405-Eng-48. P.N. also acknowledges a 
support in part by the  NSF Grant No. PHY96-05192.

\section*{Appendix}

The aim of this appendix is to understand in a simple 2D model 
how Hund's rule comes about for four-electron system. Let us assume 
that the shell spacing is fairly large, so that we may put 
two electrons in the lowest $n=0,m=0$ orbit and other two 
electrons in the next $n=0, m=1$ and $n=0, m=-1$ orbits. 
A properly 
antisymmetrized wave function is 

\begin{equation}
\Psi={1\over \sqrt{2}}[\psi_{1}(1)\psi_{-1}(2)\pm\psi_{-1}(1)\psi_{1}(2)]
|SM_S\rangle,
\label{two-elwf}
\end{equation}
where 
\begin{equation}
\psi_{1}=-{1\over \sqrt{\pi}}{\rm exp}(-{1\over 2}{\bf {\boldmath \rho}}^2)
(x+iy),\ \ \ 
\psi_{-1}={1\over \sqrt{\pi}}{\rm exp}(-{1\over 2}{\bf {\boldmath \rho}}^2)
(x-iy).
\end{equation}
The plus sign in Eq.~(\ref{two-elwf}) 
is for $S=0$ (singlet) and the minus sign is for 
$S=1$ (triplet). These two states have the orbital angular momentum 
$M=0$ but differ in their spin configurations. 
The Coulomb energies of the two states are 

\begin{equation}
\langle \Psi |{1\over |{\bf {\boldmath \rho}}_1
-{\bf {\boldmath \rho}}_2|}|\Psi \rangle
=D\pm E,
\end{equation}
where $D$ is the direct interaction energy and $E$ is 
the exchange interaction energy. Whether the 
singlet state is lower than the triplet state or not depends on the 
sign of $E$. They are given by

\begin{eqnarray}
D&=&\langle\psi_{1}(1)\psi_{-1}(2)|{1\over |{\bf {\boldmath \rho}}_1
-{\bf {\boldmath \rho}}_2|}|
\psi_{1}(1)\psi_{-1}(2)\rangle ={11\over 16}\sqrt{\frac{\pi}{2}}\nonumber \\
E&=&\langle\psi_{-1}(1)\psi_{1}(2)|{1\over |{\bf {\boldmath \rho}}_1
-{\bf {\boldmath \rho}}_2|}|
\psi_{1}(1)\psi_{-1}(2)\rangle ={3\over 16}\sqrt{\frac{\pi}{2}} > 0.
\end{eqnarray}
Since $E$ is positive, the triplet state is lower than the singlet 
state. 

When two electrons are put in $n=0, m=1$ orbit, they have the orbital 
angular momentum $M$=2 and must be in the spin singlet state 
because their spatial part is symmetric. 
The Coulomb energy of this state is given by

\begin{equation}
\langle \Psi |{1\over |{\bf {\boldmath \rho}}_1
-{\bf {\boldmath \rho}}_2|}|\Psi \rangle
=\langle\psi_{1}(1)\psi_{1}(2)|{1\over |{\bf {\boldmath \rho}}_1
-{\bf {\boldmath \rho}}_2|}|
\psi_{1}(1)\psi_{1}(2)\rangle =D,
\end{equation}  
which is equal to the average of the energies of 
the singlet and triplet states with 
$M=0$. 

In the above discussion the kinetic energy and the 
harmonic-oscillator confinement energy are not considered, but 
their contributions are the same in the three states. Also the 
Coulomb interactions between the $m=0$ and $m=\pm 1$ orbits 
give the same contribution in these three states. 
Therefore we can conclude, 
in accordance with the Hund's rule, that 
the triplet state of $(M,S)=(0,1)$ is lower than the singlet state 
of $(M,S)=(0,0)$ and also the state with $(2,0)$ is in the middle 
between them. This order is exactly the same as the one we obtained 
by the more realistic calculation (see Table II). 

\begin{figure}
\epsfbox{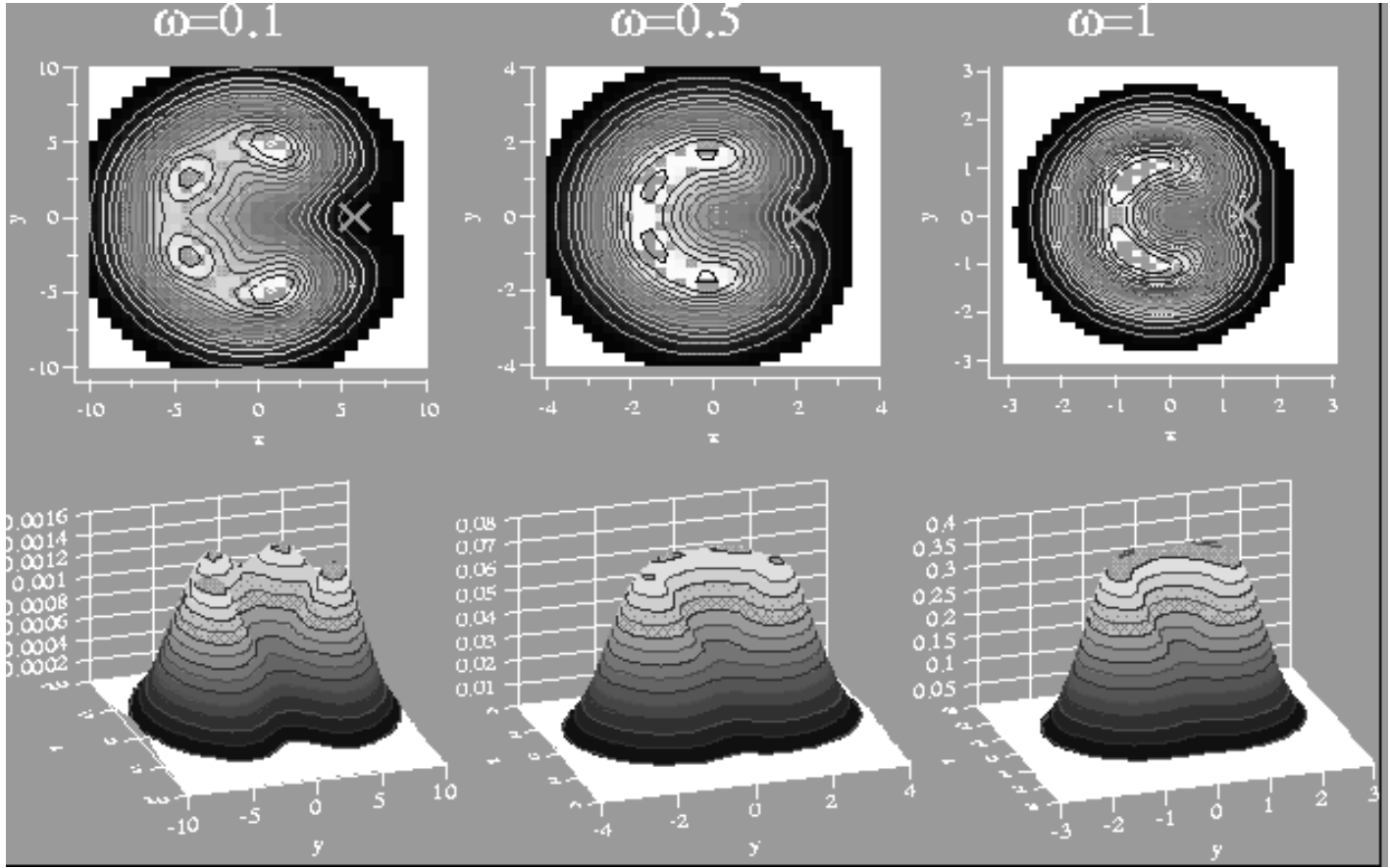}
\caption{Pair correlation function of the ground state $(M,S)=(1,1/2)$ 
of 2D five-electron system as a function of the frequency $\omega$ of the 
harmonically confining potential. The white 
cross denotes ${\bf r}_0$. Atomic units are used.}
\end{figure}

\begin{figure}
\epsfbox{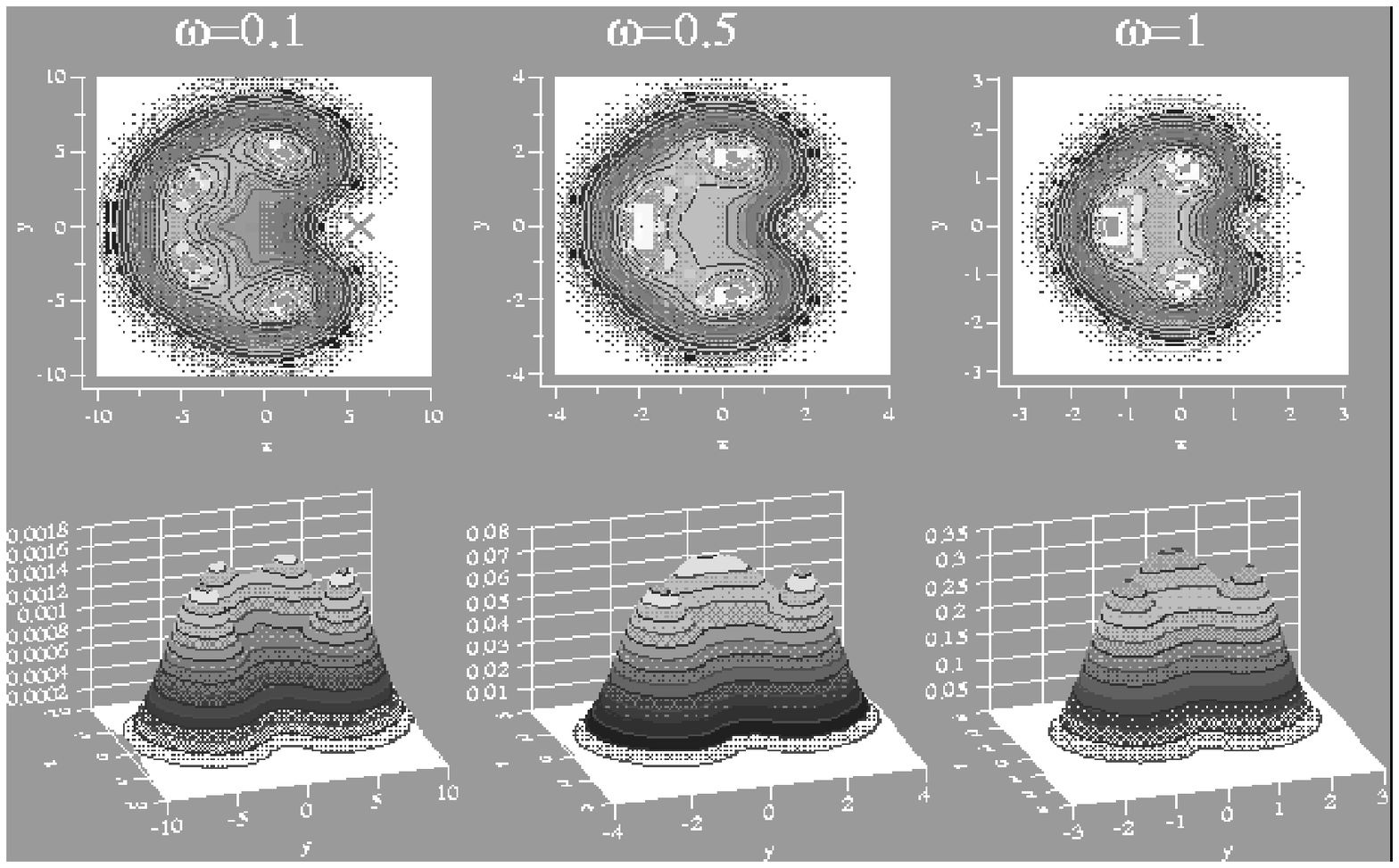}
\caption{Pair correlation fucntion of the excited state $(M,S)=(2,3/2)$ 
of 2D five-electron system as a function of the frequency $\omega$ of the 
harmonically confining potential. The white 
cross denotes ${\bf r}_0$. Atomic units are used.}
\end{figure}

\begin{figure}
\epsfbox{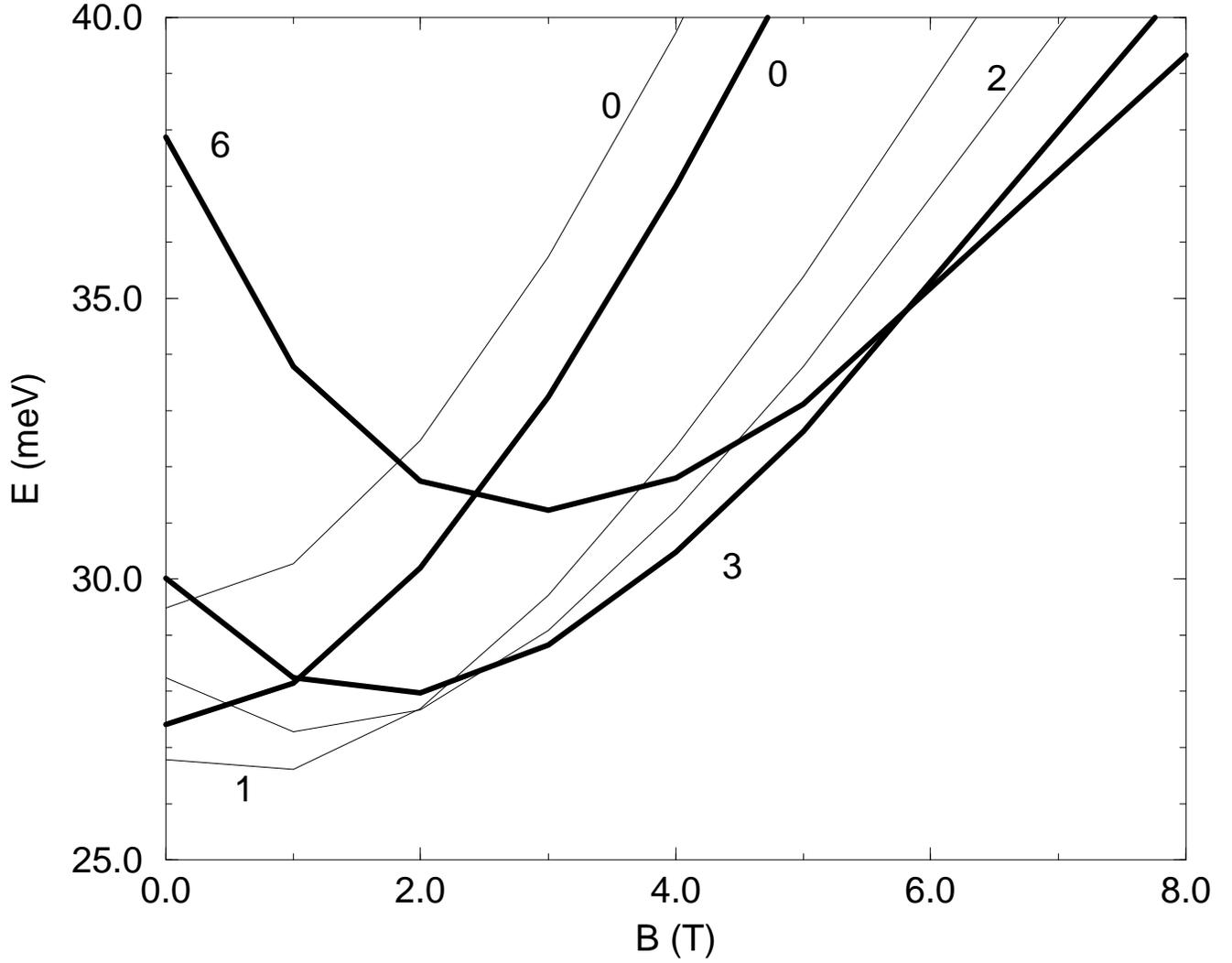}
\caption{Energies of the harmonically confined ($\hbar\omega=3.37$ meV)
lowest spin unpolarized ($S$=1/2, thin solid line) and
spin polarized ($S$=3/2 thick solid line) three-electron
states in magnetic field. The orbital angular momentum $M$ of the state 
is indicated by the number next to the curve. The Zeeman energy is 
not included.}
\end{figure}

\begin{figure}
\epsfbox{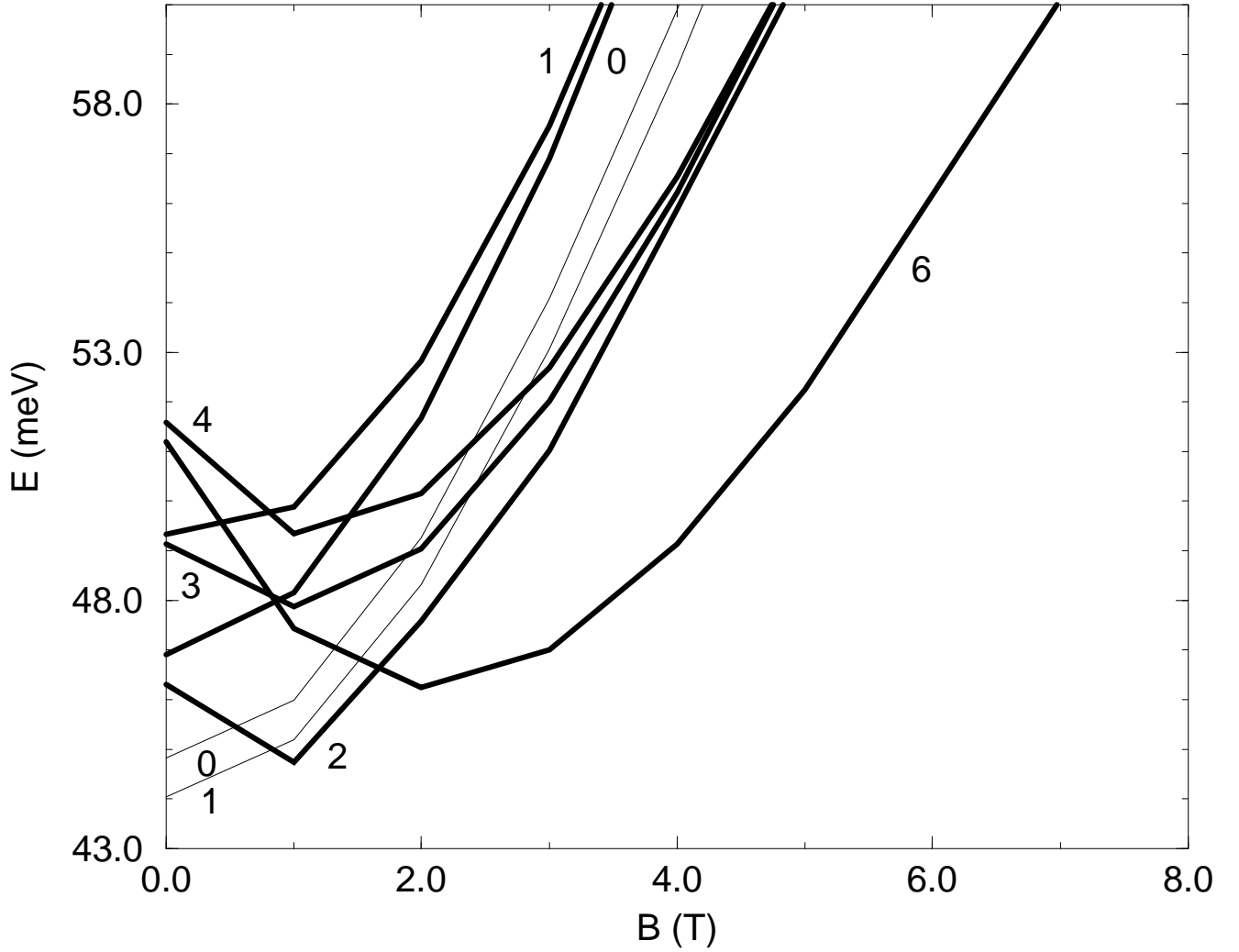}
\caption{Energies of the harmonically confined ($\hbar\omega=3.37$ meV)
lowest spin polarized ($S=$2, thick solid line) four-electron states in
magnetic field. The orbital angular momentum $M$ of the state is
indicated by the number next to the curve. The two thin solid curves
are the $S$=0 and $S$=1 (these $S$ values are indicated next
to the thin curves)
states belonging to $M$=0. The Zeeman energy is not included.}
\end{figure}

\begin{figure}
\epsfbox{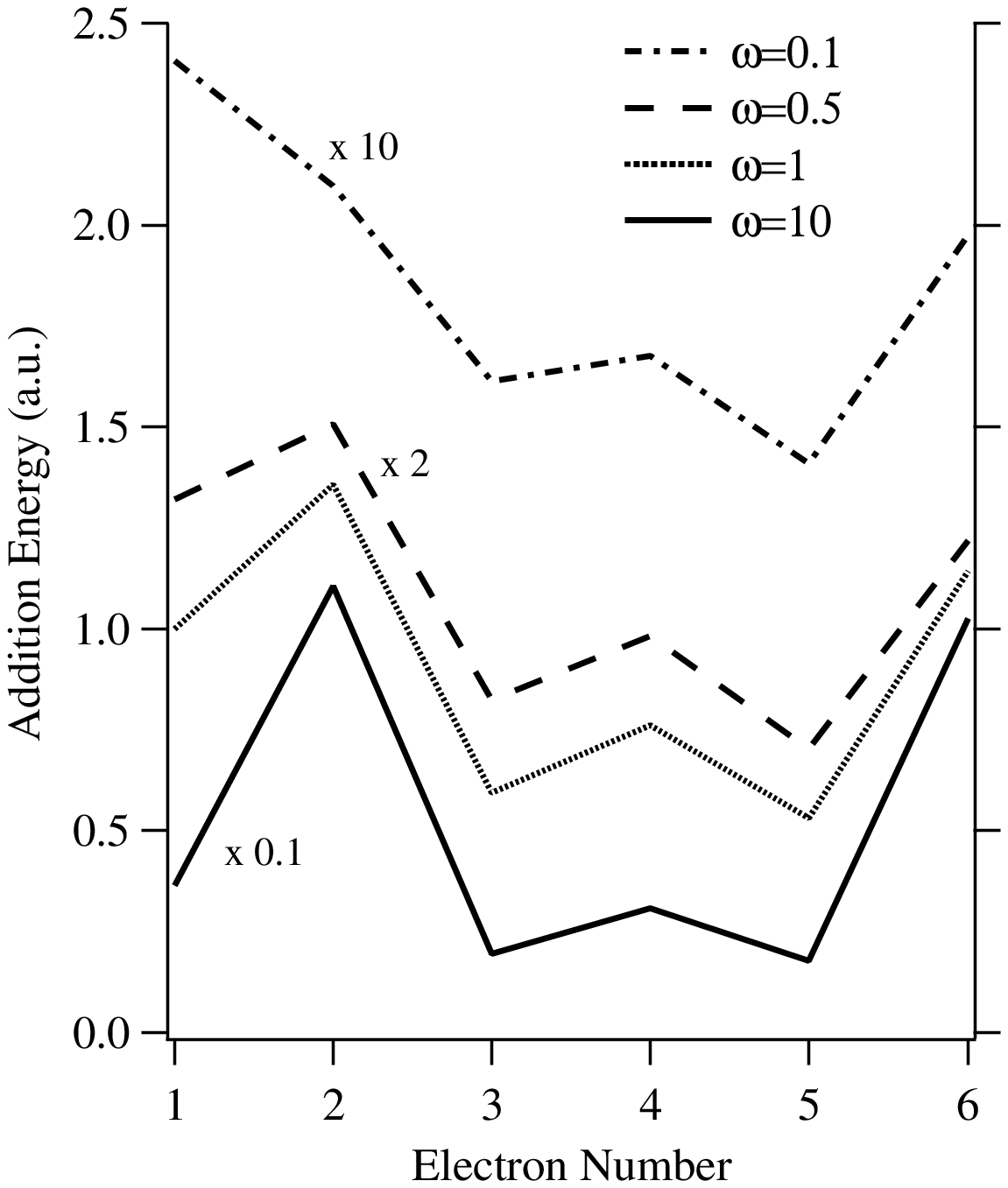}
\caption{Addition energy of harmonically confined electrons
in 2D as a function of the electron number. $\omega$ is the 
frequency of the confining potential.}
\end{figure}

\begin{figure}
\epsfbox{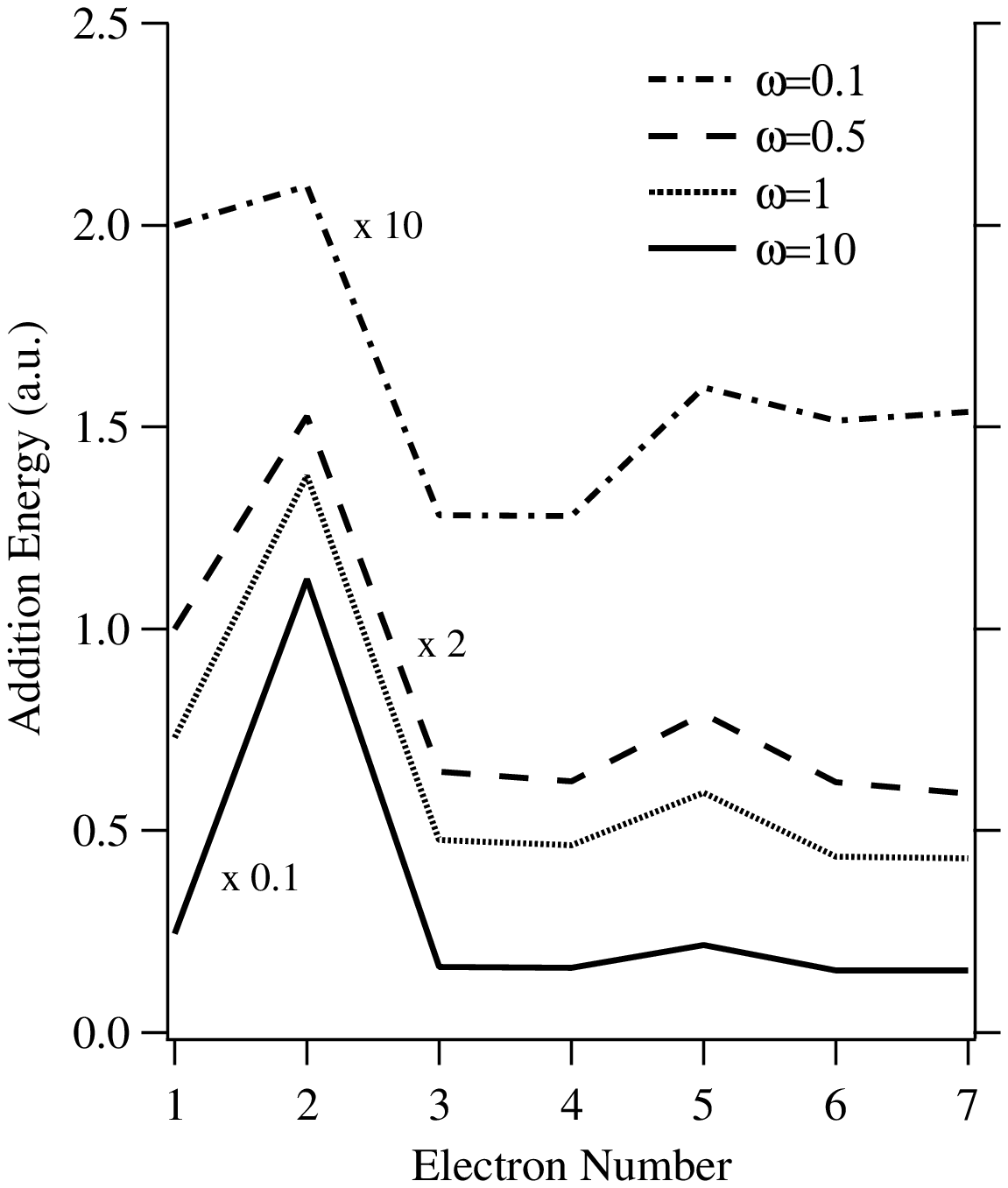}
\caption{Addition energy of harmonically confined electrons
in 3D as a function of the electron number. $\omega$ is the 
frequency of the confining potential.}
\end{figure}

\begin{figure}
\epsfbox{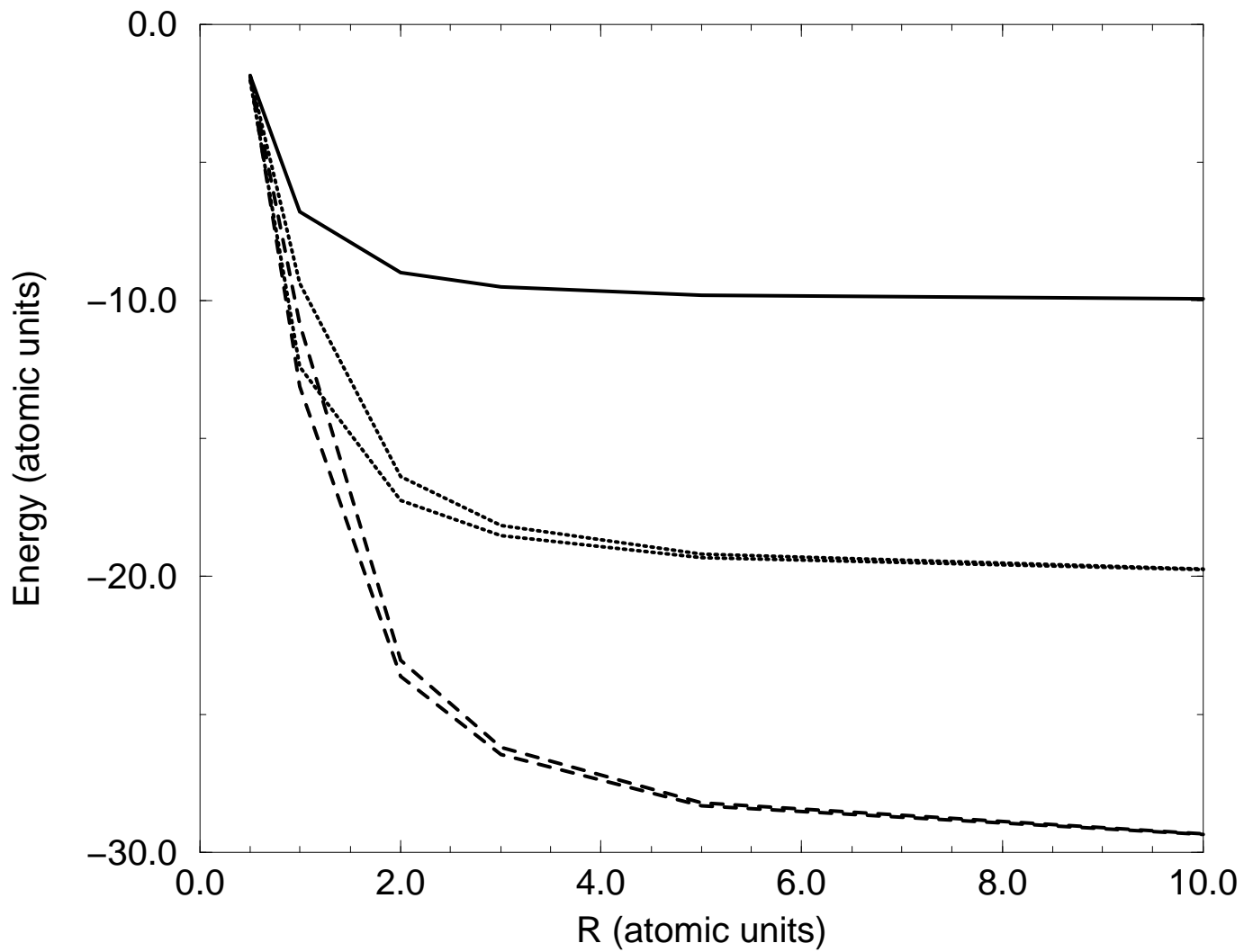}
\caption{Energies of $N_e=1$ (solid line), $N_e=2$ (dotted line) 
and $N_e=3$ (dashed line)
electron systems in a spherical quantum well 
as a function of the radius of the well. Lower dotted line: the ground 
state (0,0,+); upper dotted line: the excited state (1,1,$-$); 
lower dashed line: the ground state (1,1/2,$-$); upper dashed line: 
the first excited state (1,3/2,+).}
\end{figure}

\begin{figure}
\epsfbox{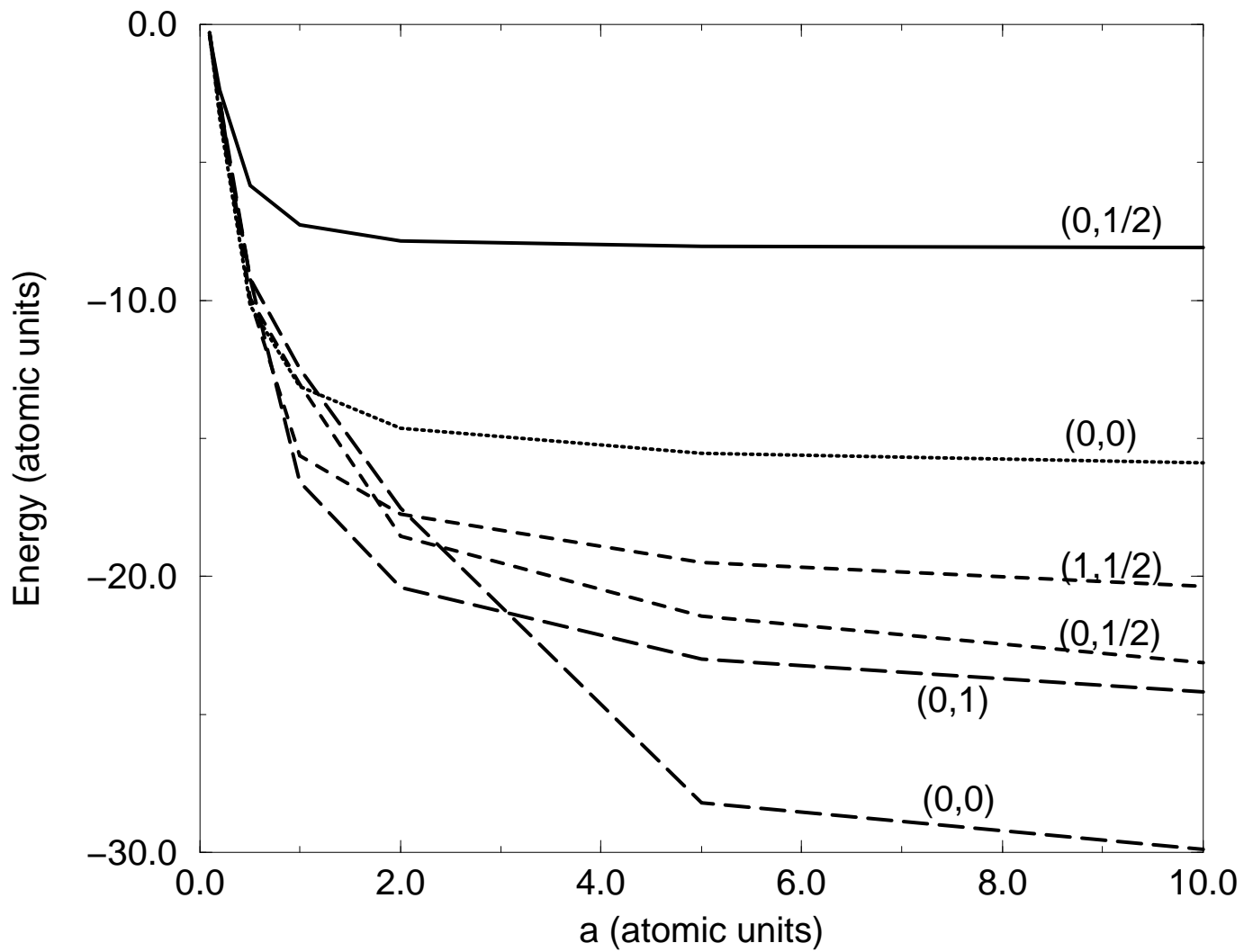}
\caption{Energy of $N_e$=1 (solid line), $N_e$=2 (dotted line), 
$N_e$=3 (dashed line)
and $N_e$=4 (long dashed line)  electron systems 
in a cylindrical quantum well 
as a function of the height of the cylinder. 
($V_0=10$ and atomic units are used.)}
\end{figure}

\begin{figure}
\epsfbox{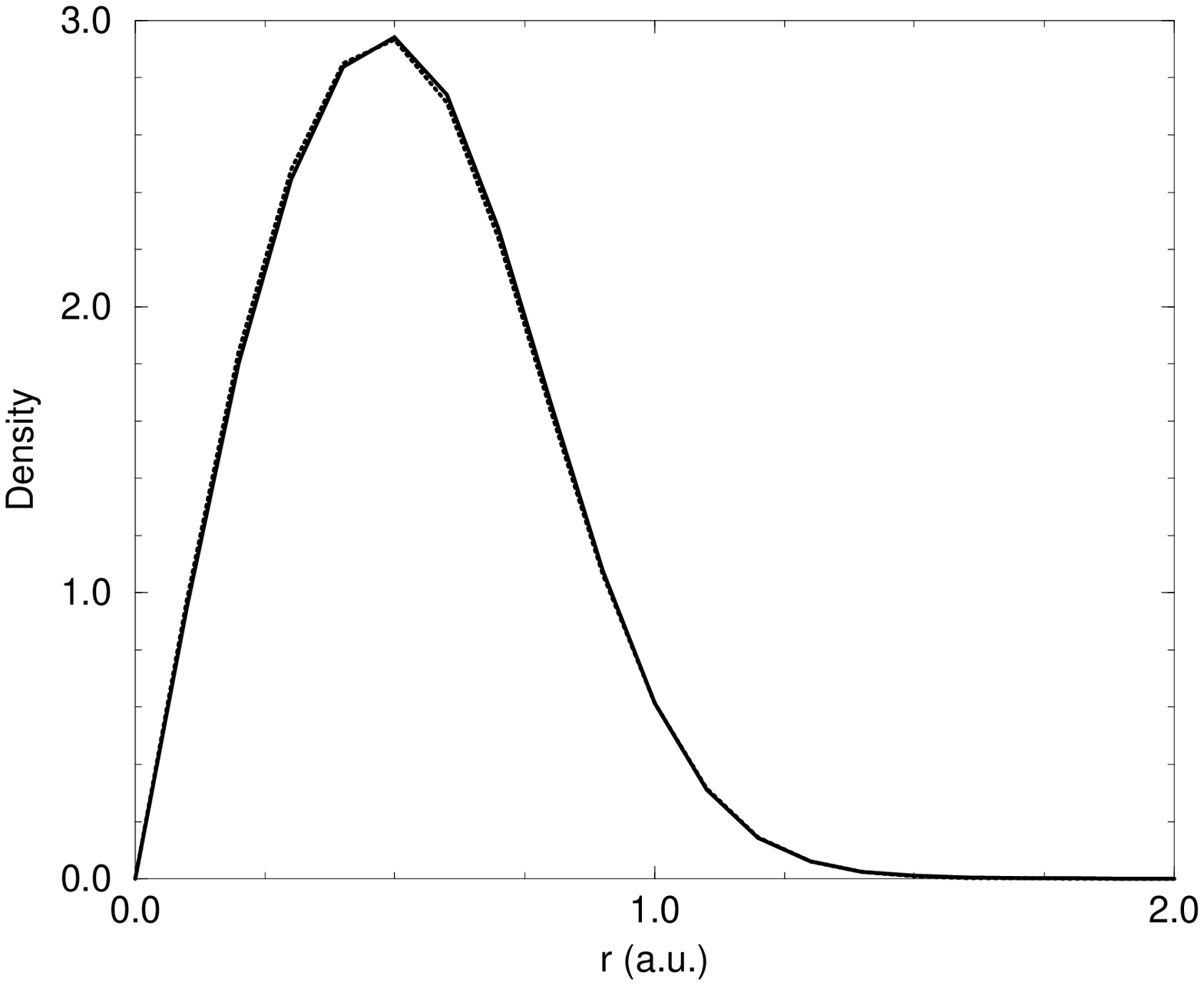}
\caption{Singlet (solid line) and triplet (dotted line) radial
density distribution of two electrons in a cylindrical
quantum dot ($a=10$).}
\end{figure}

\begin{figure}
\epsfbox{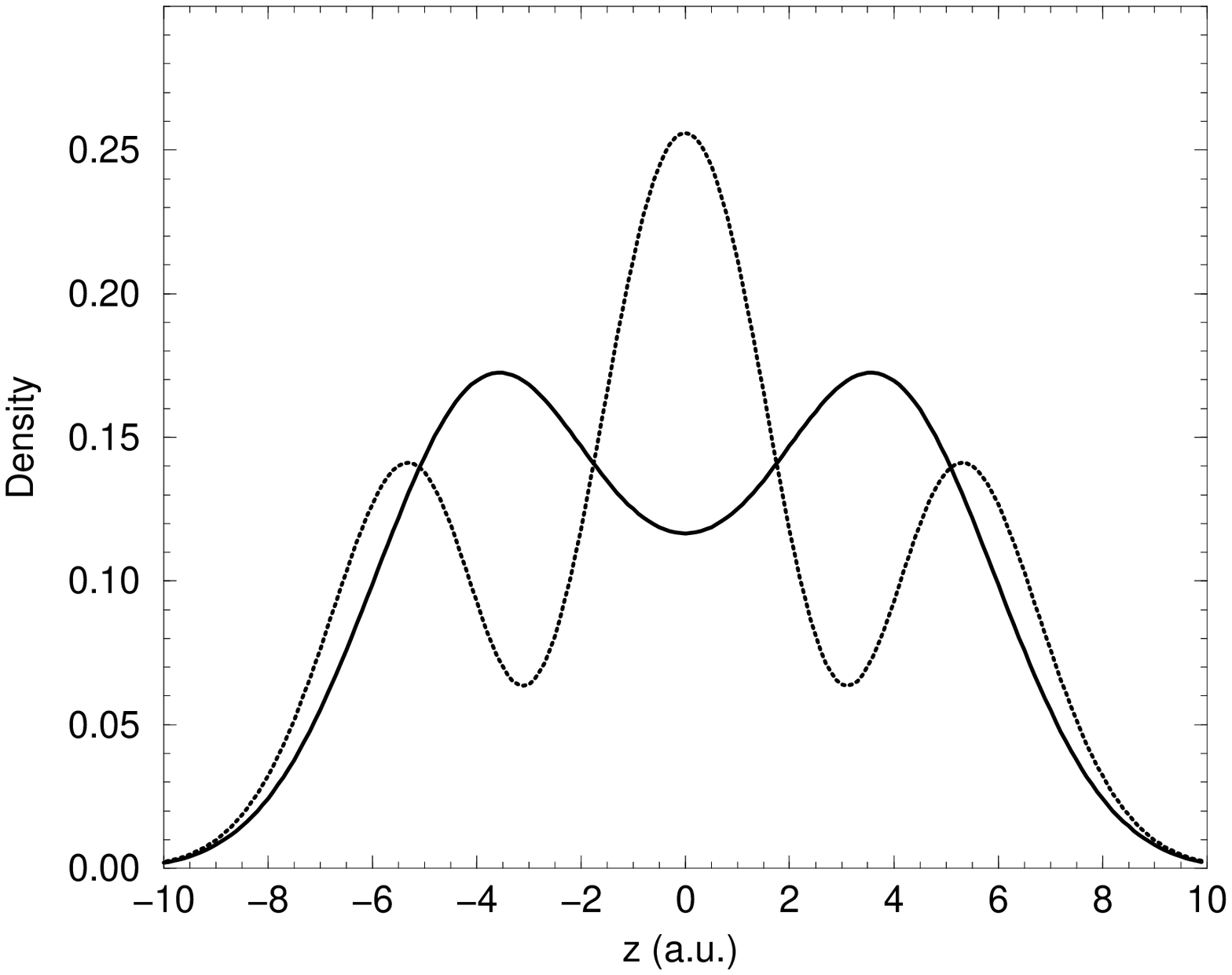}
\caption{Singlet (solid line) and triplet (dotted line)
density distribution along the $z$ direction of two electrons
in a cylindrical quantum dot ($a=10$).}
\end{figure}

\begin{figure}
\epsfbox{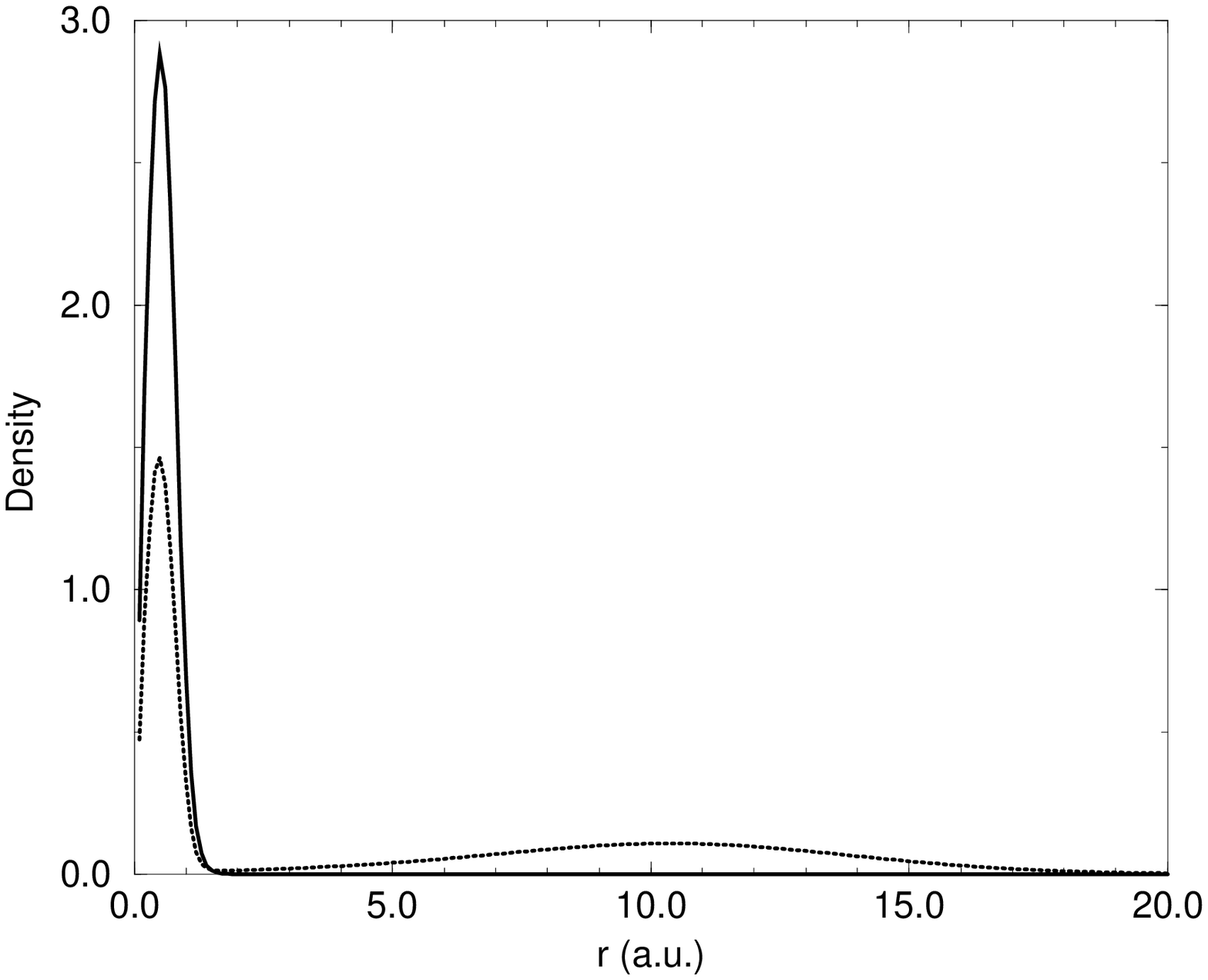}
\caption{Singlet (solid line) and triplet (dotted line) radial
density distribution of two electrons in a cylindrical
quantum dot ($a=1$).}
\end{figure}

\begin{figure}
\epsfbox{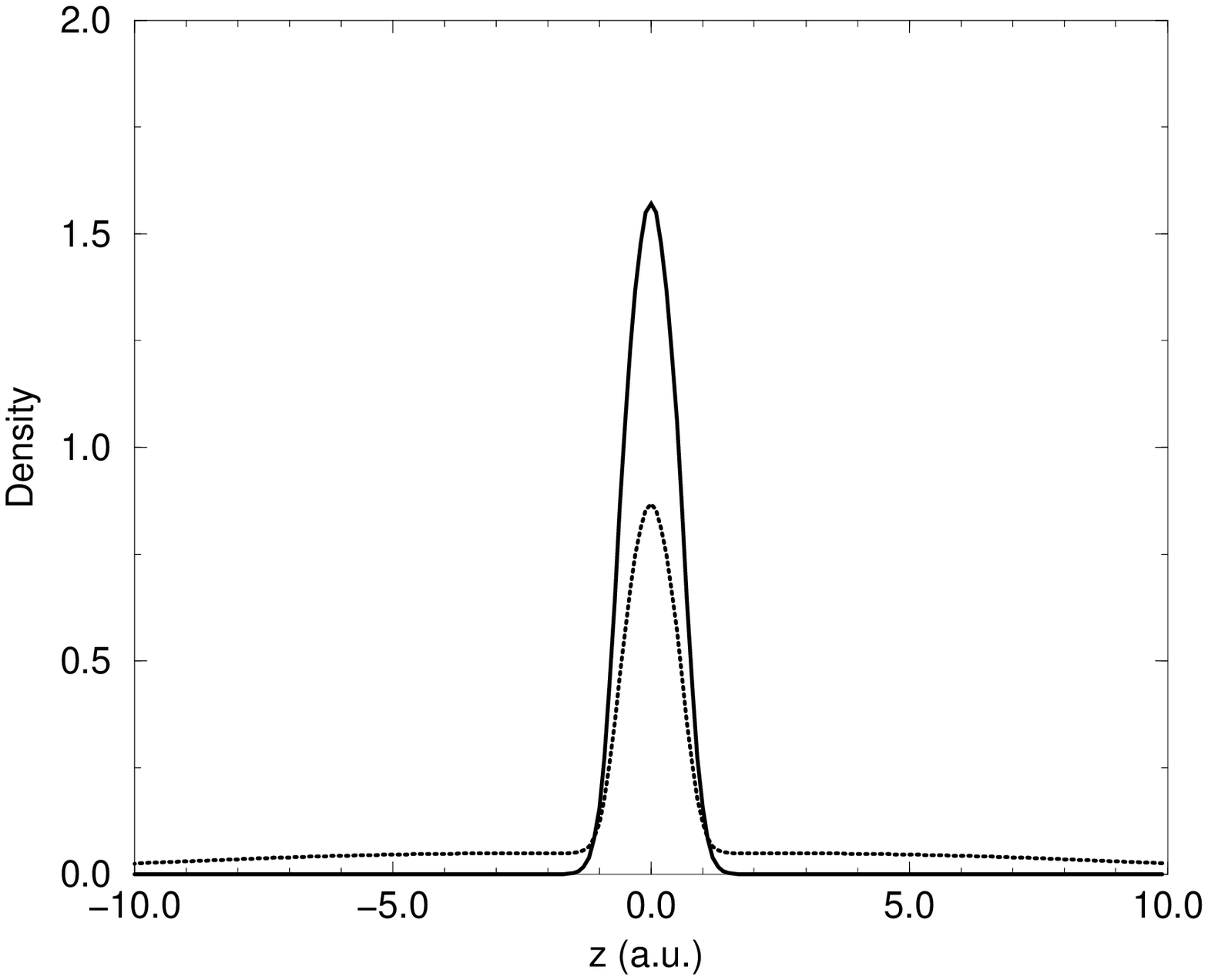}
\caption{Singlet (solid line) and triplet (dotted line)
density distribution along the $z$ direction of two electrons
in a cylindrical quantum dot ($a=1$).}
\end{figure}

\begin{figure}
\epsfbox{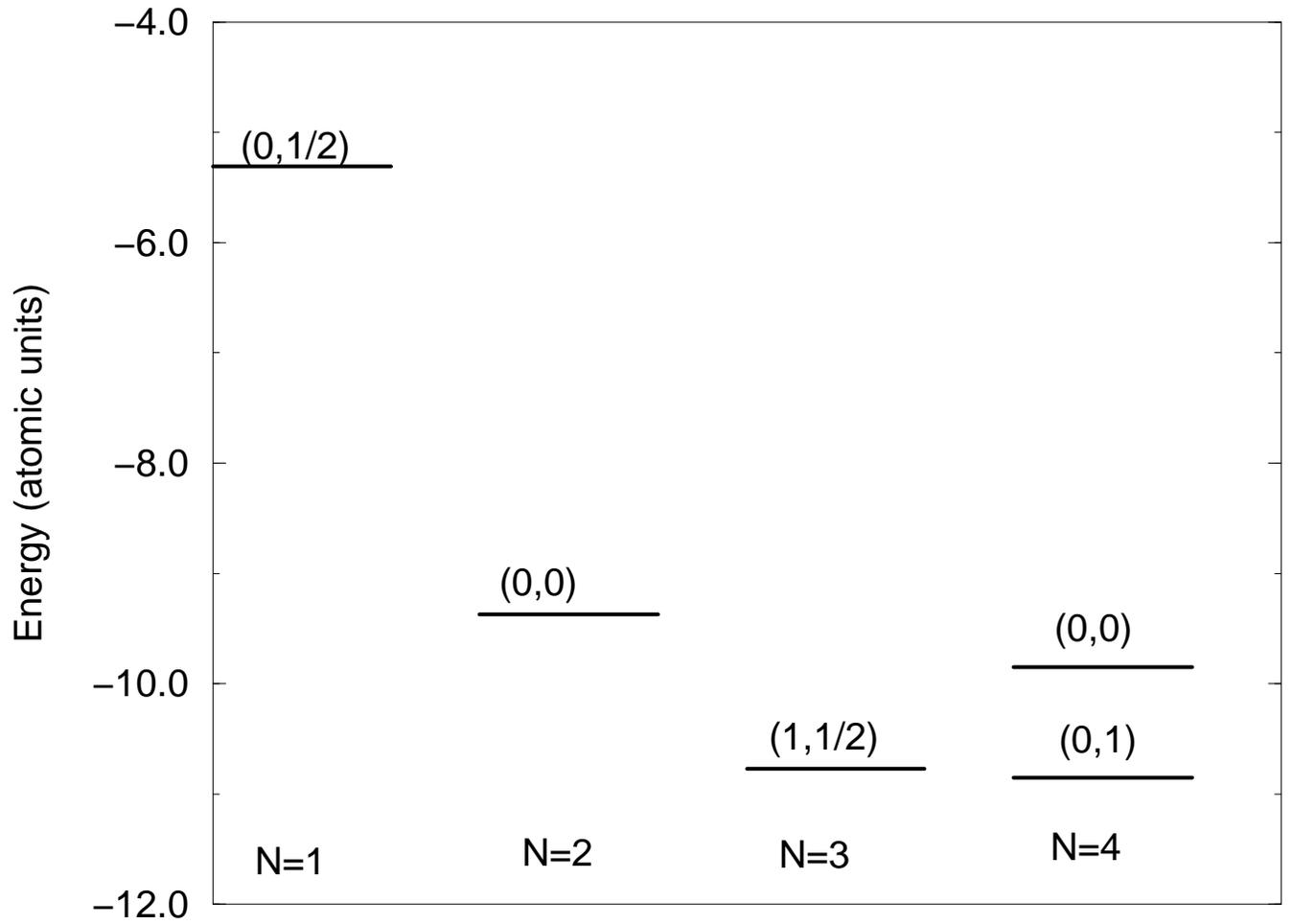}
\caption{Energy levels  of $N_e=1-4$ electron quantum rings.
The parameters of the potential are $V_0=10$, $r_1=0.5$ and $r_2=1$. 
Atomic units are used.}
\end{figure}

\begin{figure}
\epsfbox{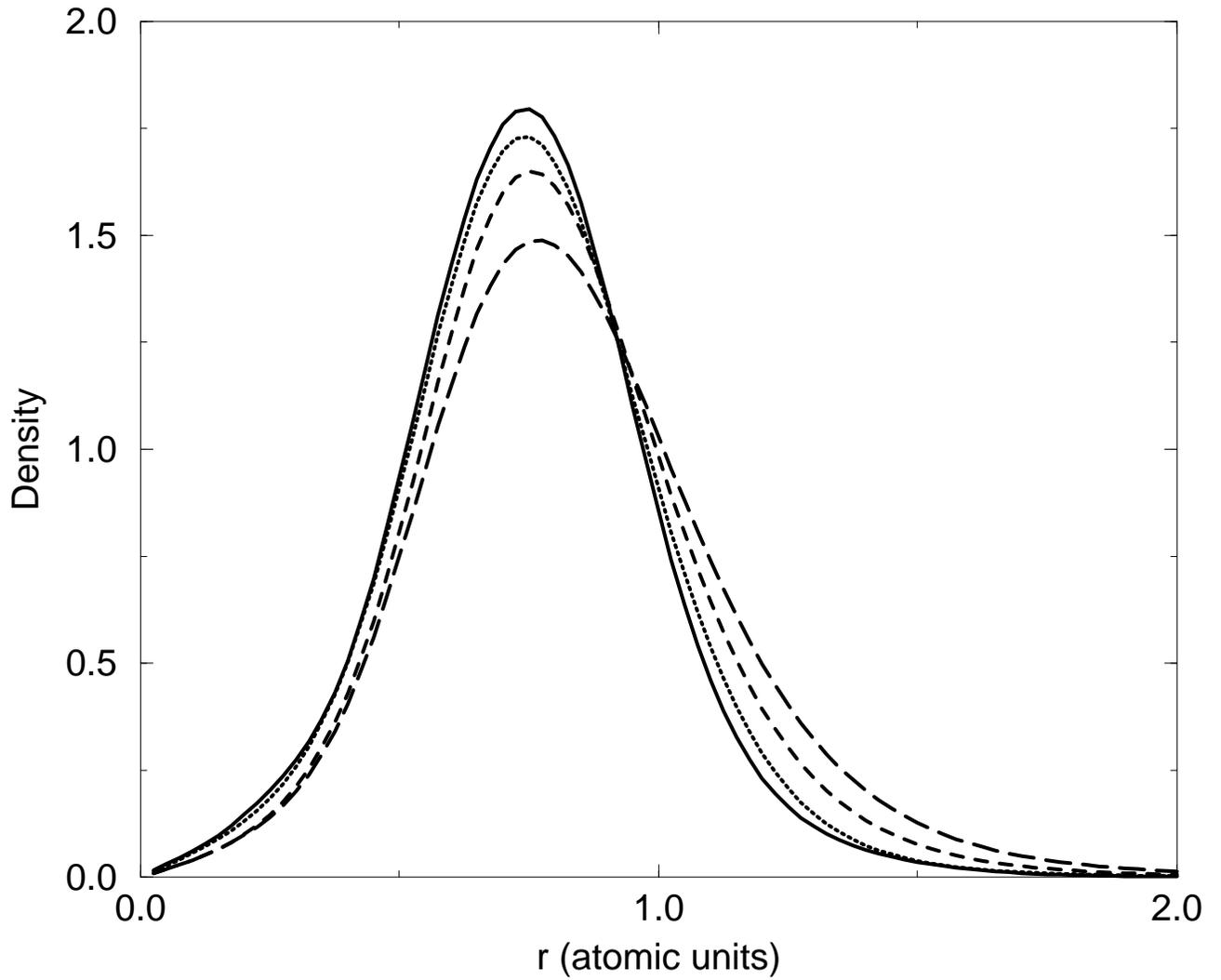}
\caption{Density distribution of $N_e=1$ (solid line), $N_e=2$ (dotted line),
$N_e=3$ (dashed line) and $N_e=4$ (long dashed line) electron quantum rings.
The parameters of the potential are $V_0=10$, $r_1=0.5$ and $r_2=1$. 
Atomic units are used.}
\end{figure}

\begin{figure}
\epsfbox{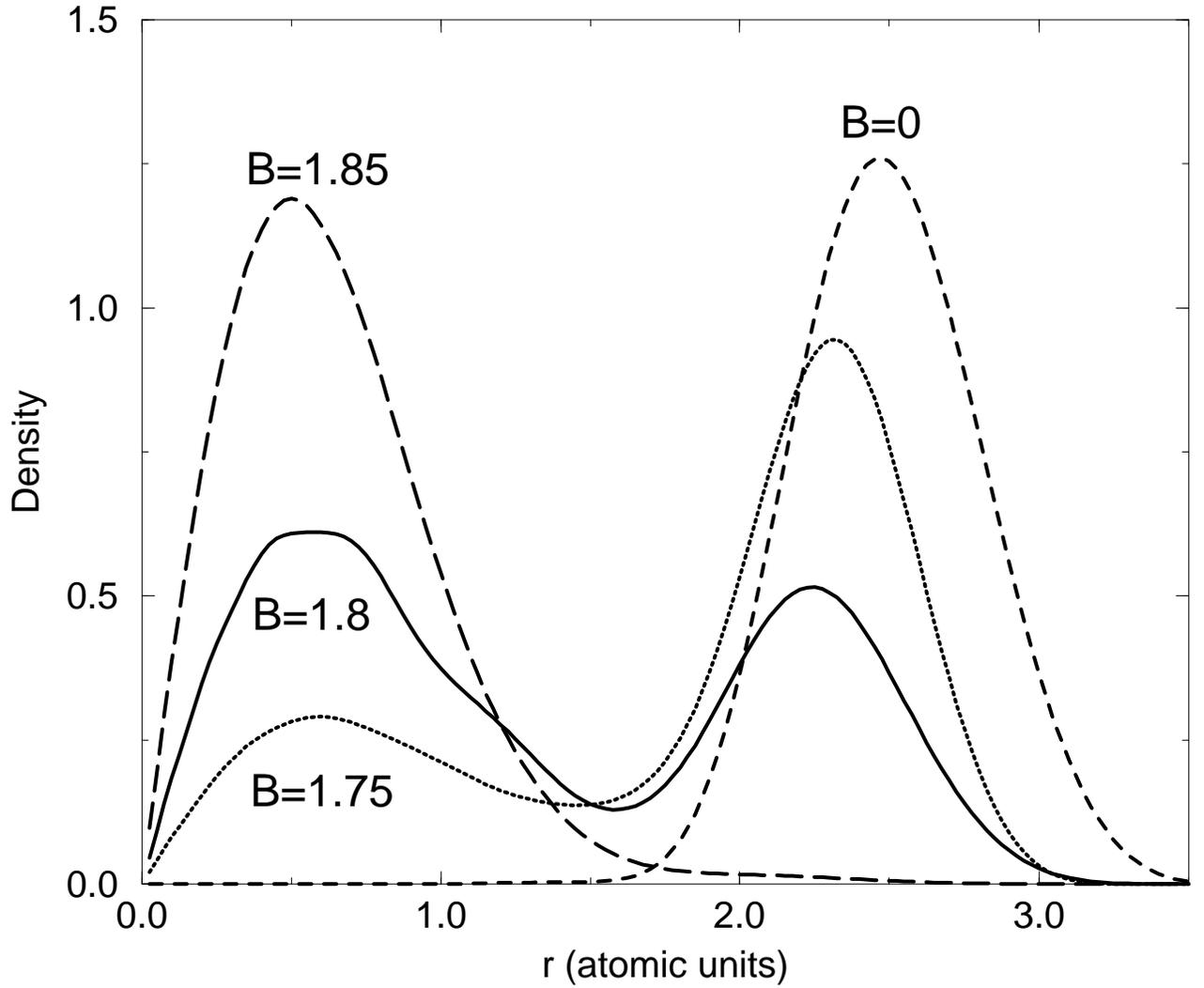}
\caption{Density distribution in a single electron quantum 
ring as a function of the magnetic field. The strength of the magnetic field 
are indicated next to the corresponding density distribution.
The parameters of the potential are $V_0=10$, $r_1=2$ and $r_2=3$. Atomic 
units are used. }
\end{figure}

\begin{table}
\caption{Comparision of the energies of harmonically confined 
2D three-electron system ($\omega=0.2841$, $\hbar \omega=3.37$ meV).
The energies are given in meV. Values in parenthesis are given in 
atomic units.}
\begin{tabular}{lllllll}
$(M,S)$ &     SVM        &  DIAG \cite{diag}&  QMC \cite{Bol96} 
                              & QMC \cite{umr} & QMC \cite{Har99}   \\
\hline
(1,1/2)&26.7827 (2.2582)&26.82&26.77&26.8214$\pm$0.0036&26.88   \\
(2,1/2)&28.2443 (2.3814)&28.27&28.30&                &28.35   \\
(3,3/2)&30.0101 (2.5304)&30.02&30.04&                &30.03   \\
\end{tabular}
\end{table}

\begin{table}
\caption{Comparision of the energies of harmonically confined 
2D electron systems ($\omega=0.28 \ \  \hbar\omega=3.32$ meV). $\eta$ is the 
virial factor.}
\begin{tabular}{lllll}
$N_e$   & $(M,S)$   &QMC\cite{umr}&   SVM    &  $\eta$      \\
\hline 
2   & (0,0)   & 1.02162(7) & 1.02164  &  0.999995    \\ 
\hline 
3   & (1,1/2) & 2.2339(3)  & 2.2320   &  0.999988    \\
\hline 
4   & (0,1)   & 3.7157(4)  & 3.7130   &  0.999971    \\
4   & (2,0)   & 3.7545(1)  & 3.7525   &  0.999982    \\                
4   & (0,0)   & 3.7135(6)  & 3.7783   &  0.999992    \\
\hline 
5   & (1,1/2) & 5.5336(3)  & 5.5310   &  0.999481    \\
\hline 
6   & (0,0)   & 7.5996(8)  & 7.6020   &  0.998912    \\
\end{tabular}
\end{table}

\begin{table}
\caption{Comparision of the energies of harmonically confined 
2D three-electron system in magnetic field($\omega=0.2841$). The energies 
are in meV except for the values in parenthesis which are 
in atomic units. $\eta$ is the virial factor.}
\begin{tabular}{lllllll}
$(M,S)$   & $B\,$(T)&  SVM    & $\eta$     &QMC\cite{Bol96}&DIAG\cite{diag}\\
\hline
(1,1/2) & 0.0 & 26.78\  (2.2582)& 0.999991&    26.77 &  26.82 \\
(1,1/2) & 1.0 & 26.61\  (2.2442)& 0.999989&    26.60 &  26.65 \\
(1,1/2) & 2.0 & 27.69\  (2.3353)& 1.000034&    27.68 &  27.74 \\
(1,1/2) & 3.0 & 29.71\  (2.5055)& 0.999987&    29.69 &  29.77 \\
(1,1/2) & 4.0 & 32.36\  (2.7283)& 1.000026&    32.32 &  32.43 \\
(1,1/2) & 5.0 & 35.39\  (2.9842)& 0.999985&    35.33 &  35.48 \\
\hline
(2,1/2) & 0.0 & 28.24\  (2.3814)& 0.999992&    28.30 &  28.27 \\
(2,1/2) & 1.0 & 27.28\  (2.2998)& 0.999925&    27.33 &  27.29 \\
(2,1/2) & 2.0 & 27.67\  (2.3338)& 0.999905&    27.72 &  27.69 \\
(2,1/2) & 3.0 & 29.09\  (2.4531)& 0.999954&    29.14 &  29.13 \\
(2,1/2) & 4.0 & 31.22\  (2.6324)& 0.999976&    31.26 &  31.26 \\
(2,1/2) & 5.0 & 33.79\  (2.8495)& 0.999963&    33.82 &  33.85 \\
\hline
(3,3/2) & 0.0 & 30.01\  (2.5304)& 0.999999&    30.04 &  30.02 \\
(3,3/2) & 1.0 & 28.24\  (2.3817)& 1.000006&    28.27 &  28.25 \\
(3,3/2) & 2.0 & 27.97\  (2.3585)& 0.999997&    28.00 &  27.98 \\
(3,3/2) & 3.0 & 28.83\  (2.4315)& 0.999999&    28.86 &  28.85 \\
(3,3/2) & 4.0 & 30.48\  (2.5703)& 0.999997&    30.51 &  30.50 \\
(3,3/2) & 5.0 & 32.63\  (2.7519)& 0.999998&    32.67 &  32.66 \\
\end{tabular}
\end{table}

\begin{table}
\caption{Energies of harmonically confined two-electron system in 3D.}
\begin{tabular}{ccccc}
$(L,S,\pi)$ & $\omega$=0.01 & $\omega$=0.5 & $\omega$=10 \\
\hline
$(0,0,+)$ & 0.07921  & 2.0000 & 32.449 \\
$(1,1,-)$ & 0.08198  & 2.3597 & 41.665 \\
$(2,0,+)$ & 0.08681  & 2.7936 & 51.338 \\
$(0,0,+)$ & 0.09696  & 2.9401 & 52.072 \\
$(3,1,-)$ & 0.09302  & 3.2538 & 61.149 \\
$(1,1,-)$ & 0.10005  & 3.3286 & 61.504 \\
\end{tabular}
\end{table}

\begin{table}
\caption{Energies of harmonically confined three-electron system in 3D.
The SVM is the stochastic variational calculation, SM is the shell model
and SM-eff is the shell model with effective interaction approach (see 
sec. II D).
Atomic units are used.}
\begin{tabular}{cccccccccc}
         & SVM   &  SM-eff   & SM   & SVM   &  SM-eff  & SM  & SVM   &  SM-eff  & SM  \\
$(L,S,\pi)$ & \multicolumn{3}{c}{$\omega$=0.01} & \multicolumn{3}{c}{$\omega$=0.5} & \multicolumn{3}{c}{$\omega$=10} \\
\hline
$(1,1/2,-)$ & 0.181936 & 0.181936 & 0.181936& 4.013240 & 4.013224& 4.013511& 61.138525 &61.138549&61.139485\\
$(1,3/2,+)$ & 0.182973 & 0.182973 & 0.182973& 4.310690 & 4.310690& 4.310712& 69.972571 &69.972571&69.972624\\
$(2,1/2,+)$ & 0.184585 & 0.184584 & 0.184584& 4.366473 & 4.366385& 4.366537& 70.315335 &70.315387&70.315871\\
$(0,1/2,+)$ & 0.191567 & 0.191568 & 0.191568& 4.467439 & 4.467459& 4.467878& 70.853077 &70.853154&70.854399\\
$(2,1/2,-)$ & 0.198201 & 0.187935 & 0.187935& 4.717817 & 4.717817& 4.717828& 79.490651 &79.490655&79.490680\\
$(1,3/2,-)$ & 0.193764 & 0.193764 & 0.193764& 4.794580 & 4.794582& 4.794614& 79.860576 &79.860582&79.860650\\
$(1,1/2,-)$ & 0.193351 & 0.193325 & 0.193325& 4.805341 & 4.797973& 4.798186& 79.890842 &79.890818&79.891346\\
$(1,1/2,-)$ & 0.199667 & 0.199656 & 0.199656& 4.960409 & 4.957257& 4.957683& 80.793524 &80.793567&80.794750\\
\end{tabular}
\end{table}

\begin{table}
\caption{Energies of harmonically confined four-electron system in 3D.
See the caption of Table V.}
\begin{tabular}{cccccccccc}
         & SVM   &  SM-eff   & SM   & SVM   &  SM-eff  & SM  & SVM   &  SM-eff  & SM  \\
$(L,S,\pi)$ & \multicolumn{3}{c}{$\omega$=0.01} & \multicolumn{3}{c}{$\omega$=0.5} & \multicolumn{3}{c}{$\omega$=10} \\
\hline
$(1,1,+)$
&0.3159&0.3141&0.3141&6.3492&6.3490&6.3502& 91.4466& 91.4459& 91.4496\\
$(2,0,+)$
&0.3177&0.3188&0.3189&6.3865&6.3865&6.3896& 91.6750& 91.6758& 91.6847\\
$(0,0,+)$
&0.3210&0.3185&0.3185&6.4462&6.4456&6.4474& 92.0260& 92.0239& 92.0297\\
$(0,2,-)$
&0.3138&0.3151&0.3151&6.5875&6.5875&6.5879& 99.9068& 99.9041& 99.9053\\
$(2,0,-)$
&0.3198&0.3181&0.3181&6.7002&6.6961&6.6980&100.5877&100.5875&100.5930\\
$(1,1,-)$ 
&0.3240&0.3195&0.3195&6.7196&6.7093&6.7105&100.6478&100.6199&100.6235\\
$(1,0,-)$ 
&0.3278&0.3251&0.3251&6.7961&6.7935&6.7963&101.0946&101.0740&101.0813\\
$(1,1,-)$ 
&0.3408&0.3232&0.3232&6.8448&6.8153&6.8169&101.3253&101.2220&101.2270\\
$(2,2,+)$ 
&0.3223&0.3212&0.3212&7.0385&7.0202&7.0205&109.5179&109.5156&109.5162\\
$(1,2,+)$ 
&0.3264&0.3313&0.3313&7.0702&7.0706&7.0719&109.7618&109.7612&109.7638\\
\end{tabular}
\end{table}
\begin{table}

\caption{Energies of harmonically confined five-electron system in 3D.
See the caption of Table V. }
\begin{tabular}{cccccccccc}
         & SVM   &  SM-eff   & SM   & SVM   &  SM-eff  & SM  & SVM   &  SM-eff  & SM  \\
$(L,S,\pi)$ & \multicolumn{3}{c}{$\omega$=0.01} & \multicolumn{3}{c}{$\omega$=0.5} & \multicolumn{3}{c}{$\omega$=10} \\
\hline
$(0,3/2,-)$ 
&0.4804&0.5141&0.5165&8.9963&8.9979&9.0032&123.357&123.3539&123.3682\\
$(2,1/2,-)$ 
&0.4858&0.5175&0.5203&9.0567&9.0526&9.0588&123.749&123.6960&123.7129\\
$(1,1/2,-)$ 
&0.4880&0.5186&0.5211&9.0954&9.0919&9.0988&123.949&123.9287&123.9482\\
$(1,3/2,+)$ 
&0.4869&0.5318&0.5359&9.3110&9.2969&9.3024&132.320&132.1385&132.1523\\
$(0,1/2,+)$
&0.4931&0.5450&0.5525&9.4443&9.4355&9.4458&133.045&132.9021&132.9265\\
$(1,3/2,+)$ 
&0.5108&0.5472&0.5537&9.7104&9.4701&9.3692&133.223&133.1205&133.1427\\
$(2,1/2,+)$ 
&0.4950&0.5357&0.5406&9.3582&9.3528&9.3599&133.471&132.487&132.5026\\
$(0,1/2,+)$ 
&0.5267&0.5561&0.5644&9.8766&9.5866&9.5990&134.204&133.8337&133.8658\\
$(2,5/2,-)$ 
&0.4829&0.5232&0.5253&9.5919&9.5891&9.5914&140.973&140.9054&140.9105\\
$(0,5/2,-)$
&0.4882&0.5306&0.5336&9.6626&9.6618&9.6648&141.270&141.2692&141.2762\\
\end{tabular}
\end{table}
\begin{table}

\caption{Energies of harmonically confined six-electron system in 3D.
See the caption of Table V.}
\begin{tabular}{cccccccccc}
         & SVM   &  SM-eff   & SM   & SVM   &  SM-eff  & SM  & SVM   &  SM-eff  & SM  \\
%         & \multicolumn{9}{c}{$\omega$} \\
$(L,S,\pi)$ & \multicolumn{3}{c}{$\omega$=0.01} & \multicolumn{3}{c}{$\omega$=0.5} & \multicolumn{3}{c}{$\omega$=10} \\
\hline
$(1,1,+)$ &0.703&0.797&0.815&12.038&12.064&12.079&157.701&157.415&157.451 \\
$(2,0,+)$ &0.743&0.801&0.819&12.080&12.101&12.118&157.910&157.643&157.681 \\
$(0,0,+)$ &0.714&0.805&0.822&12.128&12.159&12.178&158.080&157.991&158.034 \\
\end{tabular}
\end{table}

\begin{table}
\caption{Energies of harmonically confined seven-electron system in 3D.
The SM is the shell model calculation.}
\begin{tabular}{cccc}
         & SM   & SM  & SM  \\
$(L,S,\pi)$ & \multicolumn{1}{c}{$\omega$=0.01} & \multicolumn{1}{c}{$\omega$=0.5} & 
\multicolumn{1}{c}{$\omega$=10} \\
\hline
$(1,1/2,-)$ & 1.063 & 15.390 & 193.055 \\ 
$(1,3/2,-)$ & 1.084 & 15.934 & 209.998 \\
$(3,5/2,-)$ & 1.087 & 15.960 & 210.134 \\
$(0,1/2,+)$ & 1.104 & 15.672 & 201.377 \\ 
$(2,3/2,+)$ & 1.108 & 15.707 & 201.575 \\ 
$(3,3/2,+)$ & 1.113 & 15.749 & 201.802 \\ 
\end{tabular}
\end{table}

\begin{table}
\caption{Energies of harmonically confined 
eight-electron system in 3D. See the caption of Table IX.}

\begin{tabular}{cccc}
         & SM   & SM  & SM  \\
$(L,S,\pi)$ & \multicolumn{1}{c}{$\omega$=0.01} & \multicolumn{1}{c}{$\omega$=0.5} & 
\multicolumn{1}{c}{$\omega$=10} \\
\hline
$(0,0,+)$ & 1.412  & 19.038 & 230.219 \\
$(0,2,+)$ & 1.448  & 19.650 & 247.204 \\ 
$(2,1,+)$ & 1.448  & 19.653 & 247.212 \\ 
$(1,1,-)$ & 1.475  & 19.430 & 238.771 \\ 
$(3,1,-)$ & 1.479  & 19.456 & 238.915 \\
$(2,0,-)$ & 1.483  & 19.491 & 239.131 \\
$(2,1,-)$ & 1.484  & 19.496 & 239.139 \\
\end{tabular}
\end{table}

\begin{table}
\caption{Properties of harmonically confined 2D systems.}  
\begin{tabular}{ccccc}
$N_e\, (M,S)$&& $\omega=0.01$ & $\omega=0.5$&$\omega=10$\\
\hline
& $\langle H \rangle$ & 0.0738 &  1.659 & 23.652 \\
& $\langle T \rangle$ & 0.0092  &  0.443 & 9.297 \\
2 (0,0) & $\langle V_{\rm Coul} \rangle$ & 0.0369  &  0.516 & 3.372 \\
& $\langle V_{\rm con} \rangle$  &0.0277 &  0.701 & 10.983\\
& $\eta$  & 0.9999998 &0.9999995 &0.9999998 \\
\hline
& $\langle H \rangle$ & 0.176 & 3.573 & 48.365\\
& $\langle T \rangle$ & 0.016 & 0.822 & 18.286\\
3 (1,1/2) & $\langle V_{\rm Coul} \rangle$ & 0.096 &  1.286& 7.858 \\
& $\langle V_{\rm con} \rangle$  & 0.064 &  1.465 & 22.220 \\
& $\eta$  & 0.9999972 & 0.9999984& 0.9999981 \\
\hline
& $\langle H \rangle$ & 0.317 & 5.863 & 74.979  \\
& $\langle T \rangle$ & 0.018 & 1.137 & 26.836 \\
4 (0,1)  & $\langle V_{\rm Coul} \rangle$ & 0.186 & 2.391 & 14.163  \\
& $\langle V_{\rm con} \rangle$  & 0.112 & 2.335 & 33.981 \\
& $\eta$  & 0.999812 & 0.999921 & 0.999942\\
\hline
& $\langle H \rangle$ &0.515 & 8.670 & 104.642 \\
& $\langle T \rangle$ &0.0196 & 1.421 & 34.931\\
5 (1,1/2) & $\langle V_{\rm Coul} \rangle$ &0.339 & 3.874 & 23.168\\
& $\langle V_{\rm con} \rangle$  &0.159 & 3.376 & 46.543 \\
& $\eta$  & 0.9992 & 0.9995 &    0.9991  \\
\end{tabular}
\end{table}

\begin{table}
\caption{Properties of  harmonically confined 3D systems.}
\begin{tabular}{ccccc}
$N_e \, (L,S,\pi)$  & & $\omega=0.01$  & $\omega=0.5$   & $\omega=10$ \\
\hline
& $\langle H \rangle$ & 0.0792       & 2.0000   & 32.4486\\
& $\langle T \rangle$ & 0.0121       & 0.6644   & 14.4412\\
2 (0,0,+) & $\langle V_{\rm Coul} \rangle$ & 0.0366  & 0.4474  & 2.3776\\
& $\langle V_{\rm con} \rangle$  &  0.0304   & 0.8881   &15.6299\\
& $\eta$  & 0.999999      & 0.999999  & 0.999999 \\
\hline
& $\langle H \rangle$ & 0.1819        &  4.0132     & 61.1385\\
& $\langle T \rangle$ & 0.0192        &  1.1507     & 26.0867\\
3 (1,1/2,$-$) & $\langle V_{\rm Coul} \rangle$ & 0.0957  & 1.1411 & 5.9763\\
& $\langle V_{\rm con} \rangle$  & 0.0671   &  1.7214     & 29.0755\\
& $\eta$  & 0.999991       &  0.999995   & 0.999999 \\
\hline
& $\langle H \rangle$ &   0.3161    & 6.3502    & 91.446 \\
& $\langle T \rangle$ & 0.0229    & 1.5853      & 37.371 \\
4 (1,1,+) & $\langle V_{\rm Coul} \rangle$  & 0.1770  & 2.1174  & 11.132 \\
& $\langle V_{\rm con} \rangle$  & 0.1163  & 2.6475    & 42.943 \\
& $\eta$  &   0.999821   & 0.999891   & 0.999912  \\
\hline
& $\langle H \rangle$ & 0.48041  & 8.9963  & 123.36 \\
& $\langle T \rangle$ & 0.02501  & 1.9786  & 48.283 \\
5 (0,3/2,$-$)& $\langle V_{\rm Coul} \rangle$ & 0.27881 & 3.3562 & 17.808 \\
& $\langle V_{\rm con} \rangle$  & 0.17660  & 3.6615  & 57.266 \\
& $\eta$  & 0.99812      & 0.999671     & 0.999781 \\
\end{tabular}
\end{table}


\begin{thebibliography}{99}
\bibitem{ashoori} R. C. Ashoori, H. L. Stormer, J. S. Winer, 
L. N. Pfeiffer, K. W. Baldwin and K. W. West, Phys. Rev. Lett {\bf 71}
613 (1993).
\bibitem{grundmann} M. Grundmann, O. Stier, and D. Bimberg, Phys. Rev. 
{\bf B 52} 11969 (1995).
\bibitem{maksym} N. A. Bruce and P. A. Maksym, Phys. Rev. B {\bf 61}
4718  (2000).
\bibitem{diag} 
P. Hawrylak and D. Pfannkuche, Phys. Rev. Lett. {\bf 70}, 485 (1993)
\bibitem{ed} 
P.A. Maksym and T. Chakraborty, Phys. Rev. Lett. {\bf 65}, 108 (1990);
J.J. Palacios, L. Moreno, G. Chiappe, E. Louis, and C. Tejedor, 
Phys. Rev. B {\bf 50}, 5760 (1994).
\bibitem{Fuj96} M. Fujito, A. Natori, and H. Yasunaga, Phys. Rev. B 
{\bf 53}, 9952 (1996).
\bibitem{Mul96} H. M. Muller and S. Koonin, Phys. Rev. B {\bf 54}, 
14532 (1996).
\bibitem{Yan99} C. Yannouleas and U. Landman, Phys. Rev. Lett. {\bf 82}, 
5325 (1999).
\bibitem{Kos97} M. Koskinen, M. Manninen, and S.M. Reimann, Phys. Rev. Lett. 
{\bf 79}, 1389 (1997).
\bibitem{Hir99} K. Hirose and N. S. Wingreen, Phys. Rev. B {\bf 59}, 
4604 (1999)
\bibitem{Bol96} F. Bolton, Phys. Rev. B {\bf 54}, 4780 (1996).
\bibitem{Har99} A. Harju, V.A. Sverdlov, R.M. Nieminen, and V. Halonen, 
Phys. Rev. B {\bf 59} 5622 (1999).
\bibitem{umr} F. Pederiva, C. J. Umrigar and E. Lipparini, Phys. Rev B.
2000 september issue.
\bibitem{Shu99} J. Shumway, L. R. C. Fonseca, J. P. Leburton, 
R. M. Martin and D. M. Ceperley, in press.
\bibitem{dean} D. J. Dean, M. R. Strayer and J. C. Wells, 
cond-mat/9912310.
\bibitem{egger} R. Egger, W. H\"ausler, C. H. Mak, and H. Grabert, 
Phys. Rev. Lett. {\bf 82}, 3320 (1999).
\bibitem{Szaf99} B. Szafran, J. Adamowski, and S. Bednarek, 
Physica E4 1 (1999).
\bibitem{Ada00} B. Szafran, J. Adamowski, and S. Bednarek, 
Physica E 5 185 (2000).
\bibitem{Szaf00} B. Szafran, J. Adamowski, and S. Bednarek, 
Phys. Rev. {\bf B61} 1971 (2000).
\bibitem{wm1} R. Egger, W. H\"ausler, C. H. Mak, and H. Grabert,
Phys. Rev. Lett. {\bf 82}, 3320 (1999).
\bibitem{wm2} J. Harting, O. M\"ulken and P. Borrmann, cond-mat/0002269.
\bibitem{book} Y. Suzuki and K. Varga, Stochastic variational approach
to quantum mechanical few-body problems, Springer-Verlag (1998).
\bibitem{prc} K. Varga and Y. Suzuki, Phys. Rev. C{\bf 52} 2995 (1995).
\bibitem{navr1} P. Navr\'atil and B. R. Barrett, 
                Phys. Rev. C {\bf 57}, 562 (1998),
                Phys. Rev. C {\bf 59}, 1906 (1999). 
\bibitem{navr2}  P. Navr\'atil G. P. Kamuntavi\v{c}ius and B. R. Barrett,
                 Phys. Rev. C {\bf 61}, 044001 (2000).
\bibitem {VZ94} J. P. Vary and D. C. Zheng, ``The Many-Fermion-Dynamics
            Shell-Model Code'', Iowa State University (1994)
            (unpublished).
\bibitem{kvarga} K. Varga, to be published.
\bibitem{taru} S. Tarucha, D.G. Austing, T. Honda, R. J. van der Hage 
and L. P. Kouwenhoven, Phys. Rev. Lett {\bf 77} ,3613 (1996).
\bibitem{taut} M. Taut, Phys. Rev. A{\bf 48} 3561 (1993).
\bibitem{navr3} P. Navr\'atil, B. R. Barrett and W. Gl\"ockle, 
                Phys. Rev. C {\bf 59}, 611 (1999). 
\bibitem{ring1} G.E. Philipp et. al. in Diagnostic Techniques for 
Semiconductor Materials Processing II, ed. by S. W. Pang et. al. ,
MRS symposia proceedings No 406, p. 307 (1996).
\bibitem{ring2} Z. Barticevic, M. Pacheco and A. Latge, 
Phys. Rev. B{\bf 62} 6963 (2000).
\bibitem{ada} S. Bednarek, B. Szafran and J. Adamowski, Phys. Rev. 
{\bf B} 59 13036 (1999).
\end{thebibliography}
\end{document}